\documentclass[aps,showpacs,pra,amsfonts,superscriptaddress,onecolumn,notitlepage]{revtex4-1}  % for review and submission
\usepackage[utf8]{inputenc}
\usepackage[english,ngerman]{babel}
\usepackage{graphicx}  % needed for figures
\usepackage{dcolumn}   % needed for some tables
\usepackage{bm}        % for math
\usepackage{bbm}
\usepackage{amssymb}   % for math
\usepackage{amsmath}
\usepackage{mathtools}
\usepackage{changes}
\usepackage{times}
\usepackage{hyperref}
\usepackage{dsfont}
\usepackage{outlines}

\usepackage{color}
\usepackage{graphicx}%
\usepackage{dcolumn}%
\usepackage{bm}%
\usepackage{bbm}

\usepackage{amsmath}
\usepackage{amssymb}
\usepackage{mathtools}

\usepackage{changes}

\usepackage{youngtab}

\usepackage{calrsfs}
\DeclareMathAlphabet{\pazocal}{OMS}{zplm}{m}{n}

\definecolor{myurlcolor}{rgb}{0,0,0.7}
\definecolor{myrefcolor}{rgb}{0.8,0,0}
\usepackage{hyperref}
\hypersetup{colorlinks, linkcolor=myrefcolor,
citecolor=myurlcolor, urlcolor=myurlcolor}

\usepackage{cleveref}

\usepackage[final]{showlabels}

\newcommand{\bra}[1]{\langle#1|}			%
\newcommand{\ket}[1]{|#1\rangle}			%
\newcommand{\braket}[1]{\langle#1\rangle}

\newcommand*{\QED}{\hfill\ensuremath{\square}}%

\newtheorem{thm}{Theorem}
\newtheorem{cor}{Corollary}
\newtheorem{lem}{Lemma}
\newtheorem{defn}{Definition}

\begin{document}
\selectlanguage{english}

\title{The Quantum Marginal Problem for Symmetric States: Applications to Variational Optimization, Nonlocality and Self-Testing}

\date{\today}

% \author[1]{Albert Aloy}
% \author[2]{Matteo Fadel}
% \author[3]{Jordi Tura}
% \affil[1]{ICFO-Institut de Ciencies Fotoniques, The Barcelona Institute of Science and Technology, 08860 Castelldefels (Barcelona), Spain}
% \affil[2]{Department of Physics, University of Basel, Klingelbergstrasse 82, 4056 Basel, Switzerland} 
% \affil[3]{Max-Planck-Institut für Quantenoptik, Hans-Kopfermann-Straße 1, 85748 Garching, Germany}

\author{Albert Aloy}
\affiliation{ICFO-Institut de Ciencies Fotoniques, The Barcelona Institute of Science and Technology, 08860 Castelldefels (Barcelona), Spain}

\author{Matteo Fadel} 
\affiliation{Department of Physics, University of Basel, Klingelbergstrasse 82, 4056 Basel, Switzerland} 

\author{Jordi Tura} %\email{jordi.tura@mpq.mpg.de} 
\affiliation{Max-Planck-Institut f\"ur Quantenoptik, Hans-Kopfermann-Stra{\ss}e 1, 85748 Garching, Germany}

%\abstract{
\begin{abstract}
 In this paper, we present a method to solve the quantum marginal problem for symmetric $d$-level systems. The method is built upon an efficient semi-definite program that determines the compatibility conditions of an $m$-body reduced density with a global $n$-body density matrix supported on the symmetric space. We illustrate the applicability of the method in central quantum information problems with several exemplary case studies. Namely, (i) a fast variational ansatz to optimize local Hamiltonians over symmetric states, (ii) a method to optimize symmetric, few-body Bell operators over symmetric states and (iii) a set of sufficient conditions to determine which symmetric states cannot be self-tested from few-body observables. As a by-product of our findings, we also provide a generic, analytical correspondence between arbitrary superpositions of $n$-qubit Dicke states and translationally-invariant diagonal matrix product states of bond dimension $n$.
\end{abstract}
%
%}
\maketitle

\section{Introduction}

The quantum marginal problem (QMP) is ubiquitous not only in modern physics, but also in modern chemistry, where it is usually referred to as the $n$-representability problem \cite{StillingerBook1995}.
The QMP 
can be stated as 
determining whether a set of reduced density matrices (RDMs) are compatible with a global wavefunction.
The QMP arises naturally when computing physically important quantities such as the energy of a system or its entropy, as they often only depend on few particles. 
As an illustrative example, let us imagine one is interested in computing the ground energy of a $k$-local Hamiltonian $H = \sum_{i} H_i$, where each $H_i$ acts nontrivially on at most $k$ particles.
The solution to this problem is $\bra{\psi} H \ket{\psi}$, where $\ket{\psi}$ is an eigenvector of $H$ with lowest corresponding eigenvalue. Unfortunately, the amount of computational resources to describe $\ket{\psi}$ grows, in general, exponentially with the system size, rendering this approach impractical. Alternatively, one can exploit the fact that $H$ is a sum of much simpler terms, and compute instead $\bra{\psi} H \ket{\psi} = \sum_i \mathrm{Tr}[H_i \rho_i]$, where each $\rho_i$ is the reduced density matrix of $\ket{\psi}\bra{\psi}$ on the particles $H_i$ acts upon. The latter formulation, however, only appears to circumvent the exponential cost of describing $\ket{\psi}$. As a matter of fact, it actually comes at the cost of knowing the compatibility conditions of $\{\rho_i\}$ with a global state $\ket{\psi}\bra{\psi}$.

Despite this apparent simplification, the QMP has challenged the physics and chemistry communities since the 60s and every nontrivial advance has already supposed a milestone in the field \cite{KlyachkoJPConf2006}. The QMP is strongly believed to be very hard, even for a quantum computer:
It is complete for the complexity class Quantum Merlin-Arthur (QMA) \cite{LiuPRL2007} which, roughly speaking, is the analogous of NP for a quantum computer. This may be not so surprising, since the $k$-local Hamiltonian problem itself is QMA-complete \cite{KempeQIC2003}, even for $k=2$ \cite{KempeSIAM2006} or for quantum systems on a $1$D geometry \cite{AharonovCMP2009}, and existing quantum algorithms take typically exponential time to solve it \cite{Kitaev1995, PoulinPRL2009, AbramsPRL1999, GeJMP2019}.
In spite of these intractability results, tremendous progress has been achieved over the years in the QMP, even before the advent of the quantum information processing era \cite{RuskaiPR1969, ColemanBook2000, WalterScience2013}. %

A great deal of the aforementioned progress has centered onto the one-body RDM problem; i.e., determining if a set of one-body RDMs is compatible with a global (pure) state. Klyachko  showed that in this case it is sufficient to characterize the QMP compatibility conditions solely from the eigenvalues of the one-body RDMs \cite{KlyachkoJPConf2006, Klyachko2004}. For instance, the case of RDMs of a bipartite system is completely solved in terms of linear inequalities on the spectra \cite{Klyachko2004}, a rather surprising fact, since it is not obvious that compatible spectra form convex sets, let alone polytopes. This formalism has allowed for an elegant and mathematically tractable charcterization, from which further results have stemmed \cite{ChristandlCMP2014, SchillingPRA2017}, for instance, in the context of witnessing genuinely multipartite entanglement from one-body RDMs \cite{WalterScience2013}.

The few-body QMP is encompassed with substantial additional challenges. First of all, the relevant information cannot be extracted solely from the eigenvalues of the RDMs, as it was the case on the one-body case. In the one-body case, the supports of different RDMs were necessarily disjoint; therefore, the action of local unitaries did not affect global compatibility and, in consequence, only the spectrum of the one-body RDMs was relevant. In the case that one considers few-body RDMs, their supports may intersect. Therefore, these must, at least, have the same reduced density matrix on the intersection of their supports. Despite the additional requirements, some progress has been made \cite{PhDHuber}: for instance, almost all four-partite pure states are determined by their two-body marginals \cite{WyderkaPRA2017}. The QMP has also been extensively studied under the bosonic and fermionic formalism \cite{KlyachkoJPConf2006, GidofalviPRA2004, ColemanBook2000, BesteChemPhysLett2002, MazziottiPRL2012}: There, one uses the assumption of the global pure state being either fully symmetric or antisymmetric and obtains conditions on the RDMs on a given subset of parties. These RDMs are all equal due to the symmetry of the global state.

In this work we consider the following problem: Given a reduced density matrix $\sigma$ of $m$ qudits, i.e., acting on the Hilbert space $(\mathbbm{C}^d)^{\otimes m}$, is it compatible with a global reduced density matrix $\rho$ acting on the symmetric space of $n$ qudits $\mathrm{Sym}({\mathbbm C}^d)^{\otimes n}$?
Note that we denote the symmetric space as the subspace which is spanned by the Dicke states defined in \Cref{sec:Prelim}. We present the solution to this problem by analytically writing the compatibility conditions between $\sigma$ and $\rho$ and we show that they can be efficiently determined numerically as a feasibility semidefinite program (SdP) (\Cref{sec:Compat}).

{
The core results of our work can be summarized as follows:
\begin{itemize}
 \item We give the analytical conditions for any $m$-qudit RDM to be compatible with a larger $n$-qudit symmetric state
 \item We show how these compatibility conditions can be efficienly solved (polynomially in $n$, with degree $d-1$) via a SDP, thus enabling fast optimization over symmetric states
\end{itemize}
}

Our work has implications in several aspects of quantum information processing, as we show in subsequent sections with exemplary case studies. We show in \Cref{sec:Hamilt} how this provides a computationally undemanding variational approach to the ground state energy of any local Hamiltonian. We benchmark our results with different physical models, such as the Lipkin-Meshkov-Glick from nuclear physics \cite{LipkinNucPhys1965, MeshkovNucPhys1965, GlickNucPhys1965}, an Ising chain with power-law interactions, and various XXZ chains with transverse and longitudinal magnetic fields (\Cref{sec:Examples}). We further showcase how our method leads to a natural tool to optimize symmetric, few-body, Bell inequalities \cite{SciencePaper, SchmiedScience2016, AnnPhys} and we apply our method to show the ground state of some XXZ one-dimensional model with $128$ particles contains Bell correlations (\Cref{sec:Nonlocal}). In order to benchmark our results in large system sizes, we have also developed an analytical correspondence between pure symmetric states of $n$ qubits and translationally invariant diagonal matrix product states of bond dimension $n$ (\Cref{sec:MPS}) which may be of independent interest. 
Another exemplary case study that stems from our method may have implications on the self-testing of symmetric states from few-body correlators. Our method allows us to find sufficient conditions to certify which symmetric states cannot be self-tested from their marginals, by analyzing when the compatibility conditions do not lead to a unique solution (\Cref{sec:SelfTest}). 
We discuss the implications of our results and discuss further research directions in \Cref{sec:Concl}.

\section{Preliminaries}
\label{sec:Prelim}

Symmetric states constitute one of the most prominent classes of quantum states \cite{EckertAoP2002}. These are linear combinations of the so-called Dicke states, which arise naturally from the superradiance effect \cite{DickePR1954}. Dicke states and symmetric states have been successfully prepared in the laboratory in a plethora of systems, ranging from photons \cite{WieczorekPRL2009} to ultracold atoms \cite{ExpLuecke2014, Dicke3K}. Their entanglement properties have been extensively studied \cite{TuraPRA2012, AugusiakPRA2012, TuraQuantum2018}. Furthermore, Device-Independent (DI)
self-testing protocols exist for Dicke states \cite{SupicNJP2018, Fadel2017}, and symmetric states provide an advantage in quantum metrology \cite{OszmaniecPRX2016}. Symmetric states are simple to describe, as their permutational invariance allows one to circumvent the exponential growth of the Hilbert space representability problem, therefore being confined to a subspace of the multipartite Hilbert space whose dimension scales only polinomially with $n$, with degree $d-1$: More precisely, we have $\dim_{\mathbbm{C}}\mathrm{Sym}(\mathbbm{C}^d)^{\otimes n} = {n+d-1\choose d-1}$.

In the case of $n$ qubits, symmetric states are those spanned by the Dicke states, denoted $\ket{D_k^n}$
\begin{equation}
 \ket{D_k^n} \propto \sum_{\tau \in {\mathfrak S}_n} \tau(\ket{0}^{\otimes(n-k)}\ket{1}^{\otimes k}),
\label{eq:Belletti} 
\end{equation}
where $\mathfrak{S}_n$ denotes the symmetric group (the group of permutations of $n$ elements), and $\tau$ is a permutation acting on the different local Hilbert spaces.

For instance, the half-filled Dicke state of $n=4$ qubits has two excitations $k=2$, but they are delocalized among the different subsystems, in an equally-weighted coherent superposition of the same phase:
\begin{equation}
 \ket{D_2^4} = \frac{1}{\sqrt{6}}\left(\ket{0011} + \ket{0101} + \ket{0110} + \ket{1001} + \ket{1010} + \ket{1100}\right).
 \label{eq:Douglas}
\end{equation}
The number of different terms in a qubit Dicke state of $k$ excitations is given by the combinatorial expression ${n \choose k}$ and there are $n+1$ Dicke states for $d=2$.

In the general case of qudits, now one needs to specify how many $\ket{1}$ excitations there are, how many $\ket{2}$ excitations, etc. Hence, it is a natural choice to index Dicke states by partitions of $n$. We denote $\boldsymbol{\lambda} \vdash n$ as a partition of $n$ in $d$ elements, and it consists of a vector of $d$ non-negative integers that sum $d$: $\boldsymbol{\lambda} = (\lambda_0, \ldots, \lambda_{d-1})$, where $\lambda_i \in {\mathbbm{Z}_{\geq 0}}$ counts the number of excitations in state $\ket{i}$, so that $\sum_{i\in[d]}\lambda_i = n$. We will omit mentioning $d$ whenever it is clear from the context. There are ${n+d-1\choose d-1}$ such partitions and a qudit Dicke state is denoted $\ket{D_{\boldsymbol{\lambda}}}$, where
\begin{equation}
 \ket{D_{\boldsymbol{\lambda}}} \propto \sum_{\tau \in {\mathfrak S}_n}\tau(\ket{0}^{\lambda_0} \otimes \cdots \otimes \ket{d-1}^{\otimes \lambda_{d-1}}).
 \label{eq:TheObserver}
\end{equation}
The number of different terms in \Cref{eq:TheObserver} is given by the multinomial combinatorial expression
\begin{equation}
{n \choose \boldsymbol{\lambda}} = \frac{n!}{\lambda_0! \cdots \lambda_{d-1}!}.
 \label{eq:Laudrup}
\end{equation}
For the sake of a more compact notation, in the rest of this manuscript we will denote the qudit Dicke state $\ket{D_{\boldsymbol{\lambda}}}$ simply as $\ket{{\boldsymbol{\lambda}}}$, since $D$ is void of meaning.

\section{Compatibility conditions with a global symmetric state}
\label{sec:Compat}
Here we outline the fundamental building block of our work. We derive a set of necessary and sufficient conditions for compatibility of an $m$-qudit (symmetric) density matrix with a global $n$-qudit symmetric density matrix.

Let us consider a symmetric quantum system $\rho$ of $n$ qudits. Let us denote the components of $\rho$ as $\rho^{\boldsymbol{\lambda}}_{\boldsymbol{\mu}}$, where $\boldsymbol{\lambda}, \boldsymbol{\mu} \vdash n$. Thus,
\begin{equation}
 \rho = \sum_{\boldsymbol{\lambda}, \boldsymbol{\mu} \vdash n}  \rho^{\boldsymbol{\lambda}}_{\boldsymbol{\mu}} \ket{\boldsymbol{\lambda}}\bra{\boldsymbol{\mu}}.
\end{equation}

Our first goal is to compute the partial trace of $\rho$ of $n-m$ subsystems, both in the computational and in the Dicke basis. Note that since $\rho$ is symmetric, we can take the partial trace over any $n-m$ subsystems, as any choice will give the same result.

Before stating \Cref{thm:JoanGaspart}, we introduce the following notations. We define $[d] := \{0, \ldots, d-1\}$, and we shall use indices with an overhead arrow (\textit{e.g.} $\vec{i}$) to denote matrix elements in the computational basis, as opposed to indices in boldface (\textit{e.g.} $\boldsymbol{\lambda}$) that denote matrix elements in the Dicke basis. We recall once more that such boldface indeces consist in partitions of $n$, and they are associated to the Dicke basis elements through Eq.~\eqref{eq:TheObserver}.

\begin{thm}
\label{thm:JoanGaspart} 
Let $\rho$ be a symmetric state of $n$ qudits, and $m\leq n$. Let $\vec{i}, \vec{j} \in [d]^m$ and $\boldsymbol{\alpha}, \boldsymbol{\beta} \vdash m$, so that we define $\sigma:=\mathrm{Tr}_{n-m}(\rho)$; i.e.,
\begin{equation}
 \sigma = \sum_{\vec{i},\vec{j} \in [d]^m} \sigma^{\vec{i}}_{\vec{j}} \ket{\vec{i}}\bra{\vec{j}} = \sum_{\boldsymbol{\alpha}, \boldsymbol{\beta}\vdash m} \sigma^{\boldsymbol{\alpha}}_{\boldsymbol{\beta}}\ket{\boldsymbol{\alpha}}\bra{\boldsymbol{\beta}}.
\end{equation}
Then, we have the coefficients in the computational basis
\begin{equation}
 \sigma^{\vec{i}}_{\vec{j}} = \sum_{\boldsymbol{\lambda},\boldsymbol{\mu}\vdash n}\rho^{\boldsymbol{\lambda}}_{\boldsymbol{\mu}}\sum_{\boldsymbol{\kappa}\vdash n-m} \frac{{n-m\choose {\boldsymbol \kappa}}}{\sqrt{{n \choose \boldsymbol{\lambda}}{n \choose \boldsymbol{\mu}}}}\delta(\boldsymbol{w}(\vec{i})+\boldsymbol{\kappa} - \boldsymbol{\lambda})\delta(\boldsymbol{w}(\vec{j})+\boldsymbol{\kappa} - \boldsymbol{\mu})
 \label{eq:Deco}
\end{equation}
and the coefficients in the Dicke basis
\begin{equation}
 \sigma^{\boldsymbol{\alpha}}_{\boldsymbol{\beta}} = \sum_{\boldsymbol{\lambda},\boldsymbol{\mu}\vdash n}\rho^{\boldsymbol{\lambda}}_{\boldsymbol{\mu}}\sum_{\boldsymbol{\kappa}\vdash n-m} {n-m\choose {\boldsymbol \kappa}}\sqrt{\frac{{m \choose \boldsymbol{\alpha}}{m \choose \boldsymbol{\beta}}}{{n \choose \boldsymbol{\lambda}}{n \choose \boldsymbol{\mu}}}}\delta(\boldsymbol{\alpha}+\boldsymbol{\kappa} - \boldsymbol{\lambda})\delta(\boldsymbol{\beta}+\boldsymbol{\kappa} - \boldsymbol{\mu}).
 \label{eq:Cocu}
\end{equation}
\end{thm}

\paragraph{Proof of \Cref{thm:JoanGaspart}:}

We begin by noting that, in the computational basis, the partial trace of $n-m$ subsystems of an $n$-qudit density matrix $\rho$ has the following expression:
\begin{equation}
 \mathrm{Tr}_{n-m}(\rho) = \sum_{\vec{i}, \vec{j} \in [d]^m} \ket{\vec{i}}\bra{\vec{j}}\sum_{\vec{k}\in [d]^{n-m}}\rho^{\vec{i}|\vec{k}}_{\vec{j}|\vec{k}},
 \label{eq:NicolauCasaus}
\end{equation}
where the operator $|$ denotes index concatenation, and the indices have been properly rearranged so that the $n-m$ traced out parties are the last ones.

Hence, our goal is to express the symmetric state in the computational basis in order to apply \Cref{eq:NicolauCasaus} and then go back to the symmetric space. Since Dicke states are enumerated by the partitions of $m$ it will be useful to define the following function:
\begin{equation}
 \begin{array}{rccc}
 w_k:& [d]^n &\longrightarrow &\{0, \ldots, n\}\\
 & \vec{i} &\mapsto &\#\{i_p \in \vec{i}: i_p = k\}
 \end{array}
 \label{eq:Iniesta}
\end{equation}
In words, $w_k(\vec{i})$ counts how many coordinates of $\vec{i}$ are equal to $k$. It is then natural to define $\boldsymbol{w}(\vec{i}):=(w_0(\vec{i}), \ldots, w_{d-1}(\vec{i}))$. Note that, by construction, $\boldsymbol{w}(\vec{i}) \vdash n$ for every $\vec{i} \in [d]^n$.

The weight counting function $\boldsymbol w$ is useful in representing the Dicke state $\ket{\boldsymbol{\lambda}}$ in the computational basis, which we denote $\ket{D_{\boldsymbol{\lambda}}}$:
\begin{equation}
 \ket{D_{\boldsymbol{\lambda}}} = {n \choose \boldsymbol{\lambda}}^{-1/2}\sum_{\vec{i}\in [d]^n} \ket{\vec{i}} \delta(\boldsymbol{w}(\vec{i}) - \boldsymbol{\lambda}),
 \label{eq:Enhiesta}
\end{equation}
where $\delta$ is the Kronecker Delta function, which is $1$ if, and only if, its argument is the zero vector; $0$ otherwise.
Thanks to \Cref{eq:Enhiesta} we can now define the inclusion operator $\Pi:\mathrm{Sym}(\mathbbm{C}^d)^{\otimes n} \hookrightarrow ({\mathbbm{C}^d})^{\otimes n}$ onto the symmetric space
\begin{equation}
 \Pi:= \sum_{\boldsymbol{\lambda}\vdash n}\ket{D_{\boldsymbol{\lambda}}}\bra{\boldsymbol \lambda}.
 \label{eq:Pugol}
\end{equation}
We note that $\Pi^\dagger \Pi = \mathbbm{1}_{\mathrm{Sym}(\mathbbm{C}^d)^{\otimes n}}$ and that $\Pi^\dagger: ({\mathbbm{C}^d})^{\otimes n} \twoheadrightarrow \mathrm{Sym}(\mathbbm{C}^d)^{\otimes n}$ is the projector onto the symmetric space.

Let now $\rho$ be a density matrix on $\mathrm{Sym}(\mathbbm{C}^d)$, whose components are labeled $\rho^{\boldsymbol \lambda}_{\boldsymbol \mu}$ in the Dicke basis and $\rho^{\vec{i}}_{\vec{j}}$ in the computational basis. The relation between them is given by
\begin{equation}
 \rho^{\vec{i}}_{\vec{j}} = \left(\Pi (\rho^{\boldsymbol \lambda}_{\boldsymbol \mu})\Pi^\dagger\right)^{\vec{i}}_{\vec{j}} = \sum_{\boldsymbol{\lambda}, \boldsymbol{\mu}\vdash n} \delta(\boldsymbol{w}(\vec{i}) - \boldsymbol{\lambda})\delta(\boldsymbol{w}(\vec{j}) - \boldsymbol{\mu}){n \choose {\boldsymbol \lambda}}^{-1/2}{n \choose {\boldsymbol \mu}}^{-1/2} \rho^{\boldsymbol \lambda}_{\boldsymbol \mu}.
 \label{eq:ArusFonsu}
\end{equation}
Now we are ready to trace out $n-m$ parties of $\rho$, \textit{e.g.} the last. Note that $\boldsymbol{w}(\vec{a}|\vec{b}) = \boldsymbol{w}(\vec{a})+\boldsymbol{w}(\vec{b})$. Hence, using \Cref{eq:NicolauCasaus} we obtain $\sigma:=\mathrm{Tr}_{n-m}(\rho)$ as
\begin{equation}
 \sigma^{\vec{i}}_{\vec{j}} = \sum_{k\in [d]^{n-m}}\rho^{\vec{i}|\vec{k}}_{\vec{j}|\vec{k}} = \sum_{\boldsymbol{\lambda}, \boldsymbol{\mu}\vdash n}\rho^{\boldsymbol \lambda}_{\boldsymbol \mu} {n \choose {\boldsymbol \lambda}}^{-1/2}{n \choose {\boldsymbol \mu}}^{-1/2} \sum_{\vec{k}\in [d]^{n-m}}\delta(\boldsymbol{w}(\vec{i}) + \boldsymbol{w}(\vec{k}) - \boldsymbol{\lambda})\delta(\boldsymbol{w}(\vec{j})+ \boldsymbol{w}(\vec{k}) - \boldsymbol{\mu}),
 \label{eq:JoanGamper}
\end{equation}
yielding \Cref{eq:Deco}. Finally, \Cref{eq:Cocu} is obtained via the transformation $(\sigma^{\boldsymbol{\alpha}}_{\boldsymbol{\beta}})= \Pi^\dagger (\sigma^{\vec{i}}_{\vec{j}})\Pi$.
\QED

From \Cref{eq:Cocu} we can now obtain a set of necessary and sufficient conditions for a reduced density matrix $\sigma$ of $m$ qudits to be compatible with a global (possibly mixed) symmetric state of $n$ qudits, which can be expressed as the following semidefinite program (SDP):
\begin{equation}
 \begin{array}{llll}
  \min & 0&&\\
  \mbox{s.t.}& \rho &\succeq & 0\\
  &\sum_{\boldsymbol{\lambda}, \boldsymbol{\mu}\vdash n} \rho^{\boldsymbol{\lambda}}_{\boldsymbol{\mu}} a^{\boldsymbol{\lambda}, \boldsymbol{\alpha}}_{\boldsymbol{\mu}, \boldsymbol{\beta}} &=& \sigma^{\boldsymbol{\alpha}}_{\boldsymbol{\beta}} \qquad \forall \boldsymbol{\alpha}, \boldsymbol{\beta} \vdash m,
 \end{array}
 \label{eq:VictorValdes}
\end{equation}
where we have defined the coefficients $a^{\boldsymbol{\lambda}, \boldsymbol{\alpha}}_{\boldsymbol{\mu}, \boldsymbol{\beta}}$ from \Cref{eq:Cocu}; namely,
\begin{equation}
 a^{\boldsymbol{\lambda}, \boldsymbol{\alpha}}_{\boldsymbol{\mu}, \boldsymbol{\beta}} := \sum_{\boldsymbol{\kappa}\vdash n-m} {n-m\choose {\boldsymbol \kappa}}\sqrt{\frac{{m \choose \boldsymbol{\alpha}}{m \choose \boldsymbol{\beta}}}{{n \choose \boldsymbol{\lambda}}{n \choose \boldsymbol{\mu}}}}\delta(\boldsymbol{\alpha}+\boldsymbol{\kappa} - \boldsymbol{\lambda})\delta(\boldsymbol{\beta}+\boldsymbol{\kappa} - \boldsymbol{\mu}).
\end{equation}

If we think of $a^{\boldsymbol{\lambda}, \boldsymbol{\alpha}}_{\boldsymbol{\mu}, \boldsymbol{\beta}}$ as the entries of a matrix $A^{\boldsymbol \alpha}_{\boldsymbol{\beta}}$ indexed by $\boldsymbol \lambda$ and $\boldsymbol{\mu}$, then we can express the SDP \Cref{eq:VictorValdes} in canonical form as
\begin{equation}
  \begin{array}{llll}
  \min & \langle 0, \rho \rangle&&\\
  \mbox{s.t.}& \rho &\succeq & 0\\
  &\langle A^{\boldsymbol \alpha}_{\boldsymbol{\beta}}, \rho \rangle &=& \sigma^{\boldsymbol{\alpha}}_{\boldsymbol{\beta}} \qquad \forall \boldsymbol{\alpha}, \boldsymbol{\beta} \vdash m,
 \end{array}
 \label{eq:AlbertJorquera}
\end{equation}
where $\langle \cdot, \cdot \rangle$ denotes the Hilbert-Schmidt inner product.

If the SDP \Cref{eq:AlbertJorquera} is feasible, then $\sigma$ admits an extension to a symmetric Dicke state of $n$ qudits $\rho$ which is precisely the solution of that SDP. In \Cref{sec:SelfTest} we discuss about the uniqueness of that solution.

\subsection{Variational ansatz}
\label{sec:Hamilt} 
We can now easily modify the SDP \Cref{eq:AlbertJorquera} to optimize any linear functional $H$ on $\rho$, while maintaining compatibility over a given marginal state $\sigma$. This is done by considering the SDP
\begin{equation}
  \begin{array}{llll}
  \min & \langle H, \rho \rangle&&\\
  \mbox{s.t.}& \rho &\succeq & 0\\
  &\langle A^{\boldsymbol \alpha}_{\boldsymbol{\beta}}, \rho \rangle &=& \sigma^{\boldsymbol{\alpha}}_{\boldsymbol{\beta}} \qquad \forall \boldsymbol{\alpha}, \boldsymbol{\beta} \vdash m.
 \end{array}
 \label{eq:Piquembauer}
\end{equation}

The most interesting case arises when such a functional can be expressed as a sum of terms with support on, at most, $m$ qudits. This includes many cases of physical interest, such as Hamiltonians or Bell operators composed of, at most, $m$-body interactions/correlators. In this case, let us denote $H= \sum_{i} H_i$. Then, $\langle H, \rho \rangle$ can be expressed as a linear combination of terms of the form $\langle H_i, \sigma \rangle$, namely:
\begin{equation}
  \begin{array}{llll}
  \min & \sum_i \langle H_i, \sigma \rangle&&\\
  \mbox{s.t.}& \rho &\succeq & 0\\
  &\langle A^{\boldsymbol \alpha}_{\boldsymbol{\beta}}, \rho \rangle &=& \sigma^{\boldsymbol{\alpha}}_{\boldsymbol{\beta}} \qquad \forall \boldsymbol{\alpha}, \boldsymbol{\beta} \vdash m.
 \end{array}
 \label{eq:Zampaburguers}
\end{equation}
Note that in \Cref{eq:Zampaburguers}, both $\rho$ and $\sigma$ are treated as positive-semidefinite variables. The positive-semidefiniteness of $\sigma$ is automatically implied by that of $\rho$. In fact, $\sigma$ can be completely removed from \Cref{eq:Zampaburguers} and embedded into the objective function; however we keep it in this form for clarity of exposition. The form of \Cref{eq:Zampaburguers} is thus useful to optimize a functional $H$ that depends only on the marginal information contained in the reduced states, while keeping compatibility with a global symmetric state. Recall that the size of $\rho$ depends polinomially on $n$, with a degree $d-1$, so that this procedure is efficient for systems of qudits of large $n$ and fixed $d$.

\section{Some Applications}
The aim of this section is to illustrate several applications in various, apparently uncorrelated, problems in quantum information, all of which have deep roots in the QMP.

In \Cref{sec:Examples} we apply \Cref{eq:Zampaburguers} to benchmark our method as a variational ansatz to find a fast, upper bound, to the ground state energy and, in some cases, to well approximate the ground state of several paradigmatic Hamiltonians. In \Cref{sec:Nonlocal} we adapt our method to optimize Bell functionals composed of symmetric, few-body observables. In \Cref{sec:MPS} we provide a method to approximate any $n$-qubit Dicke state with a translationally-invariant (TI) diagonal matrix product state (MPS) of bond dimension $n$. Finally, in \Cref{sec:SelfTest} we show how our method can be used to show which symmetric states cannot be self-tested from few-body marginals.

\subsection{Benchmarking the variational ansatz}
\label{sec:Examples}

Here we consider some Hamiltonians of exemplary spin models, in order to benchmark the performance of the variational ansatz presented in \Cref{sec:Hamilt}. Furthermore, we provide in \Cref{sec:time} the runtime and estimated resources consumed by our method, compared to DMRG as a benchmark.

The variational ansatz approximates the ground state and energy of an $m$-local Hamiltonian by an $m$-qudit RDM (denoted $m$-RDM for short) compatible with a global many-body symmetric state. We expect the variational ansatz to provide a good approximation when the coupling interactions are similar between all pairs and to provide exact results when the ground state lies in the symmetric space.

\subsubsection{Lipkin-Meshkov-Glick model}
\label{subsec:LMG}
Let us start by considering a spin model for which our variational method recovers the ground state exactly. We picked as an example the Lipkin-Meshkov-Glick (LMG) model \cite{LipkinNucPhys1965, MeshkovNucPhys1965, GlickNucPhys1965}, which involves long-range interactions that result in ground states that are symmetric under any permutation of the particles. The LMG model was originally proposed in nuclear physics to describe phase transitions in nuclei. However, nowadays it also serves to describe \textit{e.g.} two-mode Bose-Einstein condensates experiments, since it captures the physics of interacting bosons in a double-well trapping potential. Furthermore, in its isotropic version, the ground states of the LMG model are pure Dicke states, which have been shown to display Bell correlations \cite{SciencePaper, AnnPhys, TuraPRX2017}. The phase transitions of the general model are also well understood \cite{LMG-JILA}.
Exact solutions are known \cite{PanPhysLettB1999, LinksJPhysA2003, RibeiroPRE2008}, which we use here to benchmark our method.

In particular, we consider the following LMG Hamiltonian which describes a set of $n$ spin$-1/2$ particles with anisotropic long-range interactions under an external transverse magnetic field $h$:

\begin{equation}
\pazocal{H} = -\frac{\lambda}{n}\sum\limits_{i < j} \left( \sigma_x^{(i)} \sigma_x^{(j)} + \gamma \sigma_y^{(i)}\sigma_y^{(j)}\right)-h\sum\limits_{i=1}^n \sigma^{(i)}_z,
\label{eq:LMG}
\end{equation}
where $\sigma^{(i)}_k$ denotes the Pauli matrix in position $i$ and direction $k$, $\lambda >0$ correspond to ferromagnetic (FM) interactions, $\lambda <0$ correspond to anti-ferromagnetic (AFM) interactions, and $\gamma$ marks the anisotropy in the coupling terms, with $\gamma=1$ being the isotropic case.

By adapting the SDP optimization problem in \Cref{eq:Zampaburguers}, we can use the 2-RDM compatibility constraints to approximate its ground state by means of SDP. In particular, the optimization problem to find the best approximation within the variational ansatz can be reduced to the following SDP:

\begin{equation}
  \begin{array}{llll}
  \min &  \mathrm{Tr} \left(\tilde{\pazocal{H}} \sigma\right) &&\\
  \mbox{s.t.}& \rho &\succeq & 0\\
  &\langle A^{\boldsymbol \alpha}_{\boldsymbol{\beta}}, \rho \rangle &=& \sigma^{\boldsymbol{\alpha}}_{\boldsymbol{\beta}} \qquad \forall \boldsymbol{\alpha}, \boldsymbol{\beta} \vdash m,
 \end{array}
\label{eq:SDPvm}
\end{equation}
where we emphasize that $\tilde{\pazocal{H}}:=-{n\choose 2}\frac{\lambda}{n}\left(\sigma_x\otimes \sigma_x+\gamma\sigma_y\otimes \sigma_y\right)-nh\left(\sigma_z\otimes\mathbbm{1}+\mathbbm{1}\otimes \sigma_z\right)/2$ is now an effective 2-body Hamiltonian.

\begin{figure}[h!]
\centering
\begin{minipage}{.33\textwidth}
  \centering
  \includegraphics[width=1\linewidth]{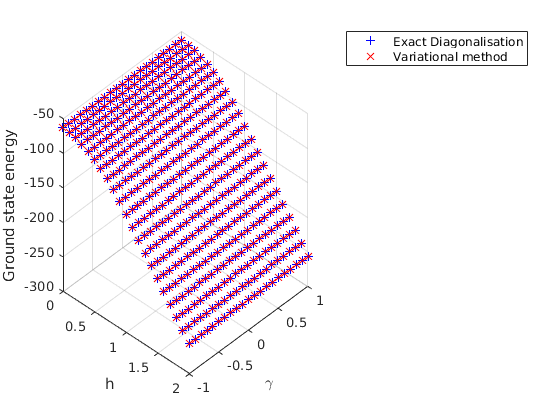}
\end{minipage}%
\begin{minipage}{.33\textwidth}
  \centering
  \includegraphics[width=1\linewidth]{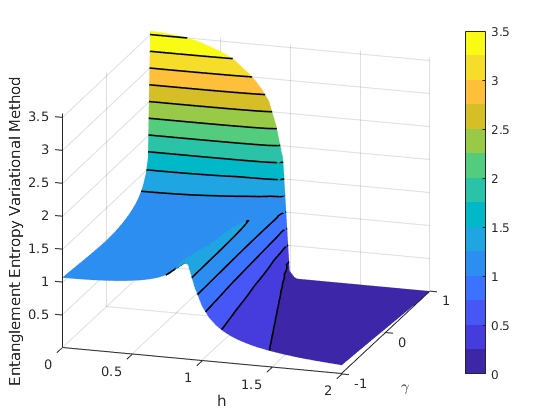}
\end{minipage}
\begin{minipage}{.33\textwidth}
  \centering
  \includegraphics[width=1\linewidth]{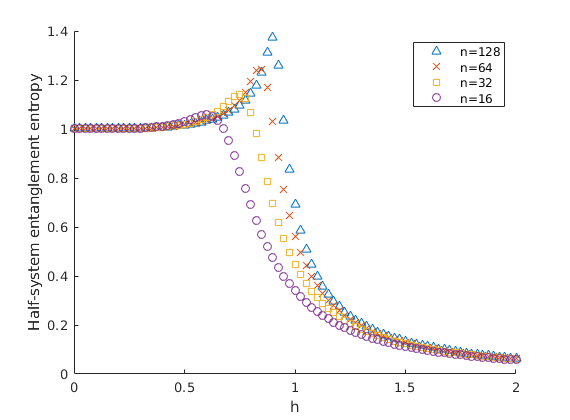}
\end{minipage}
\caption{Numerical results for the LMG Hamiltonian \eqref{eq:LMG} with $h=\lambda=1$. Left: The variational method faithfully reproduces the ground state energy of the LMG model for $n=128$. Up to numerical error ($\sim 10^{-14}$ using SeDuMi \cite{SeDuMi}) the values coincide.
Center: Always for $n=128$, we use the RDM compatibility constraints to obtain the half-system RDM from the ground state found by the variational method. This enables us to compute the entanglement entropy, and to characterize the phase diagram of the model \cite{LMG-JILA}. Right: Half-system entanglement entropy for $\gamma=0$ and different values of $n$ obtained from the variational method using the RDM compatibility constraints in the symmetric basis. For $\gamma\neq 1$, one can appreciate the anomaly in $h=1$ as we increase $n$, which becomes a critical point in the asymptotic limit.
}
\label{fig:LMG}
\end{figure}

In \Cref{fig:LMG} we show that the ground state energy of the model we considered is faithfully recovered using our variational method.

We note that this method unlocks the possibility to have access to an efficient description of the $n$-qubit state (given in the symmetric basis \Cref{eq:Cocu}), since it is a matrix of size $(m+1) \times (m+1)$. This, in turn, allows to obtain any associated $m$-RDM $\forall 1 \leq m \leq n$. This enables us to study the method against extensive quantities such as entropy, since it is now easy to obtain the $m$-block size entanglement entropy $\pazocal{S}_{m,n}=-\sum_{i=0}^{m}p_i\log_{2}p_i$ by finding the eigenvalues $p_i$ of the $m$-RDM \cite{LMG-JILA}. In \Cref{fig:LMG} we have used such a procedure to obtain the half-system entanglement entropy. The method reproduces the features of the LMG model phase diagram, as expected.

\subsubsection{Ising chain with variable-range interactions under a transverse field}
\label{sec:IsingPowerLaw}
As a second example, we consider the Ising model with variable-range interactions in a transverse field. The tuneable interaction range allows us to explore how our variational method performs as the range of interactions decreases from the infinite-range case (equivalent to the LMG model) to the nearest-neighbour case. In particular, we consider the following Hamiltonian for an Ising chain with decaying power-law interactions:
\begin{equation}
\pazocal{H} = \sin(\theta)\sum\limits_{i < j} J_{ij}\sigma_z^{(i)} \sigma_z^{(j)}+ \cos(\theta)\sum\limits_{i=1}^n \sigma^{(i)}_x,
\label{eq:IsingChainPowerLaw}
\end{equation}
where $J_{ij}=|i-j|^{-\alpha}$, the paramether $\alpha$ tunes the range of interactions, and $\theta >0$ ($\theta <0$) results in AFM (FM) interactions.  We note that in the limit $\alpha \rightarrow 0$ all pairs interact with the same strength \cite{FadelQuantum2018}, whereas in the other extreme $\alpha \rightarrow \infty$ we have interactions only between nearest neighbours. The phase diagram for this model has been extensively characterised \cite{KoffelPRL2012, KnapPRL2013, GabbrielliNJP2019}. In particular, for $\alpha>0$ the model exhibits three phases: an ordered ferromagnetic phase for $-\pi/2\leq \theta < \theta_c^-(\alpha)$; a disordered paramagnetic phase for $\theta_c^-(\alpha)< \theta < \theta_c^+(\alpha)$; and an ordered anti-ferromagnetic phase for $  \theta_c^+(\alpha)< \theta\leq \pi/2$. Notably, such a model has also been shown to display Bell correlations at the critical points for the ferromagnetic couplings \cite{PigaPRL2019}.

In order to construct the variational method for \Cref{eq:IsingChainPowerLaw}, we consider the SDP in \Cref{eq:SDPvm} with the effective Hamiltonian $\tilde{\pazocal{H}}:= J \sin(\theta)\sigma_z\otimes \sigma_z + n\cos(\theta)(\sigma_x\otimes  \mathbbm{1}+\mathbbm{1}\otimes \sigma_x)/2$, where $J:=\sum\limits_{i< j}J_{ij}$.

\begin{figure}[h!]
\centering
\begin{minipage}{.5\textwidth}
  \centering
  \includegraphics[width=0.8\linewidth]{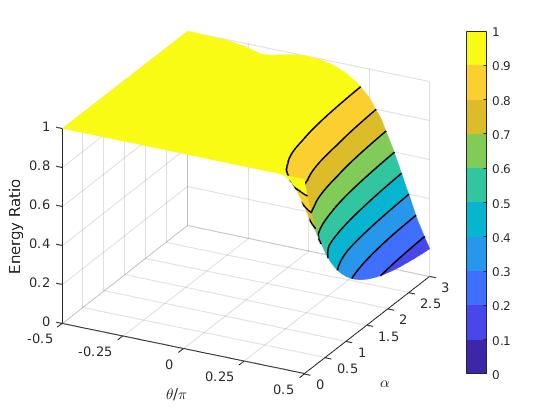}
\end{minipage}%
\begin{minipage}{.5\textwidth}
  \centering
  \includegraphics[width=0.8\linewidth]{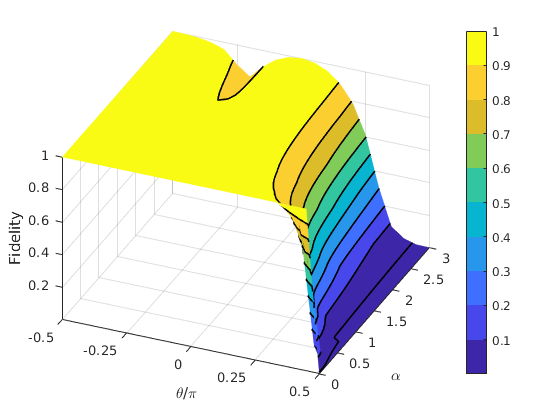}
\end{minipage}
\caption{Numerical results for the Ising Hamiltonian \eqref{eq:IsingChainPowerLaw} with $n=10$, compared to exact diagonalization. Left: energy ratio, Right: ground state fidelity. As expected, the case $\alpha=0$ is faithfully recovered by the variational method. However, we observe that as the value of $\alpha$ increases (the range of interaction decreases) the variational method fails to capture the anti-ferromagnetic phase for which the fidelity eventually drops to zero. The ground state energy and fidelity in the ferromagnetic and paramagnetic phases are well approximated, although for large values of $\alpha$ there is a little discrepancy near the critical points $\theta \approx \theta_c^-(\alpha)$. }
\label{fig:IsingChainPowerLaw_EnergyFidelity}
\end{figure}

In \Cref{fig:IsingChainPowerLaw_EnergyFidelity} we compare the ground states obtained using our variational method (VM) with those obtained from exact diagonalisation (ED), in terms of relative energy and fidelity. To compare the ground state energies we look at their ratio, $E_0^{\text{VM}}/E_0^{\text{ED}}$. For this, a few comments are in order: First of all, note that the ground state energy is negative and sufficiently far from zero to constitute a good approximation of $1-\delta$, where $\delta$ is the relative error.
Second, we have chosen the ratio as a figure of merit, instead of the relative error, 
as it gives a better visual comparison with the fidelity also plotted in \Cref{fig:IsingChainPowerLaw_EnergyFidelity}. To compute the ground state fidelity we use the definition $F(\rho_{\text{ED}},\rho_{\text{VM}}):=\left(\rm{Tr}\sqrt{\sqrt{\rho_{\text{ED}}}\rho_{\text{VM}}\sqrt{\rho_{\text{ED}}}}\right)^2$.
For this model we expect that the analytical solutions for the long-range interaction regime are well approximated by the one for the LMG model, where our variational method yields exact results. However, in the transition from $\alpha \gg 1$ to $\alpha \simeq 0$ the quality of the approximation is \textit{a priori} not so clear. In \Cref{fig:IsingChainPowerLaw_EnergyFidelity} it can be appreciated that for $\alpha>0$ the method fails to capture the anti-ferromagnetic phase $\theta_c^+(\alpha)< \theta$, eventually yielding fidelities close to zero. On the other hand, for values $\theta<\theta_c^+(\alpha)$ the method provides a good approximation, even though for sufficiently large $\alpha$ the method is less accurate near the critical point $\theta_c^-$.

\begin{figure}[h!]
\centering
\begin{minipage}{.49\textwidth}
  \centering
  \includegraphics[width=0.8\linewidth]{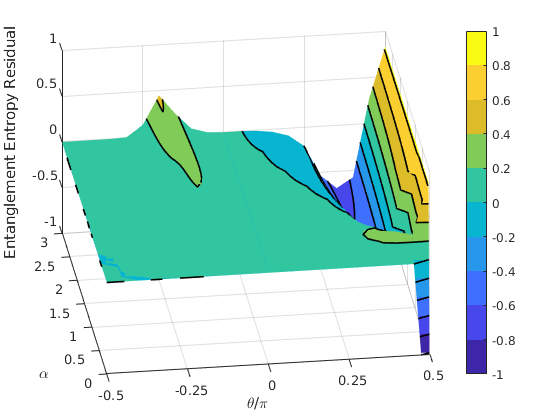}
\end{minipage}
\begin{minipage}{.49\textwidth}
  \centering
  \includegraphics[width=0.8\linewidth]{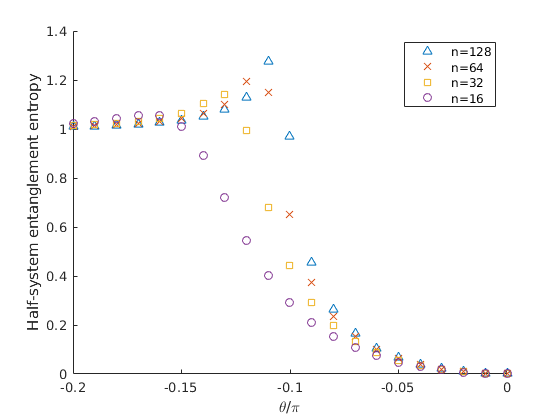}
\end{minipage}
\caption{Numerical results for the Ising Hamiltonian \eqref{eq:IsingChainPowerLaw} with $n=10$. Left: Half-system entanglement entropy residual obtained from comparing VM with ED (a value of $0$ indicates an exact result). Although the paramagnetic to anti-ferromagnetic phase transition is not reproduced by the VM, one observes that the transition from ferromagnetic to paramagnetic is well approximated by looking at the discrepancies for values of $\alpha \gtrsim 1.5$. Right: Half-system entanglement entropy for $\alpha = 2$. Note that the transition from ferromagnetic to paramagnetic manifest itself in the vicinity of $\theta \simeq -0.1 \pi$ for $n=10$. From this observation we conjecture that our VM can be used to extrapolate some critical points where phase transitions occurs. We remark that the ED has not been used in this case and that the behavior observed arises from the VM alone.}
\label{fig:IsingChainPowerLaw_hsEE}
\end{figure}

In \Cref{fig:IsingChainPowerLaw_hsEE} we compare half-system entanglement entropies obtained from the VM and ED. As expected, we observe a discrepancy in the anti-ferromagnetic phase, because of the little overlap between the ground state and the symmetric space in that region.
Interestingly, we see from \Cref{fig:IsingChainPowerLaw_hsEE} that in the ferromagnetic regime one can use the VM (without ED) to approximate its phase transition (between paramagnetic and ferromagnetic) for different number of particles, and extrapolate its asymptotic limit. Such an approximation naturally works better as the range of interactions increases.

The VM has the potential to identify a phase transition in the model due to the following argument: One might expect that in many cases of interest, and for a finite number of particles, the sudden change in nature (\textit{e.g.} symmetry) of the ground state is actually ``smeared'' rather smoothly around the critical point. Therefore, everywhere around this point the ground state may still have some overlap with the symmetric space, which is the one considered by our variational method. The numerical evidence presented in \Cref{fig:IsingChainPowerLaw_hsEE} supports this conjecture.

\subsubsection{Ising chain for nearest neighbours interactions}
\label{sec:IsingNN}
So far, in \Cref{subsec:LMG} we focused on the extreme case of infinite-range interactions (the LMG), and in \Cref{sec:IsingPowerLaw} we explored how our variational method behaves as we decrease the range of interactions. Let us now investigate what happens in the other extreme case: an Ising model with nearest-neighbours interactions in a transverse field. The Hamiltonian we consider is:

\begin{figure}[h!]
\begin{minipage}{.33\textwidth}
  \centering
  \includegraphics[width=1\linewidth]{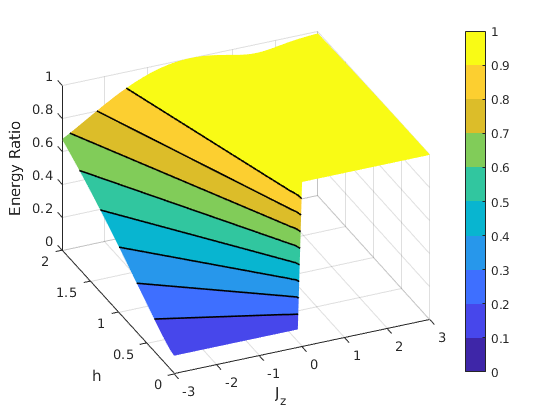}
\end{minipage}
\begin{minipage}{.33\textwidth}
	\centering
	\includegraphics[width=1\linewidth]{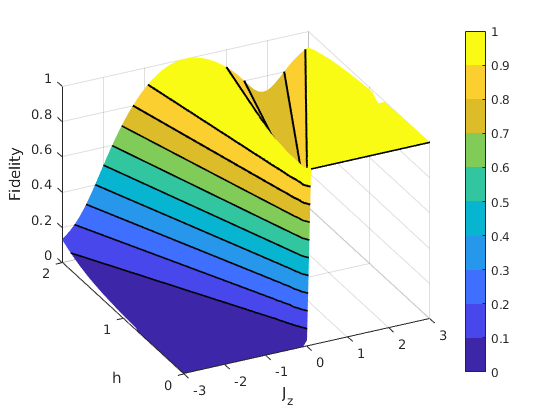}
\end{minipage}
\begin{minipage}{.33\textwidth}
	\centering
	\includegraphics[width=1\linewidth]{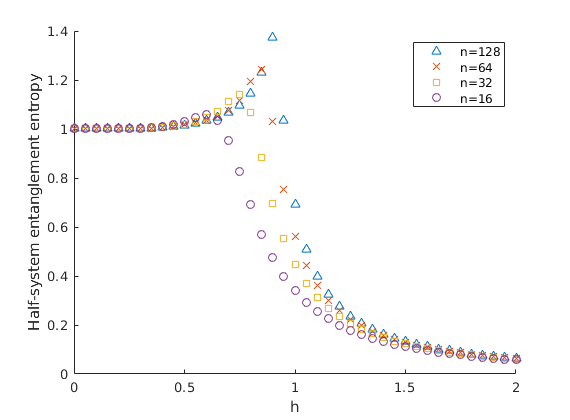}
\end{minipage}
\caption{Numerical results for the Ising Hamiltonian \eqref{eq:IsingChainNN} with $n=10$. Left and center: comparison of the ground state energy and fidelity with respect to results from exact diagonalization. We observe that energies disagree mostly in the AFM case, due to its inherent asymmetry. The FM case is in general well approximated, with only some minor discrepancies near the values that hint at a critical point in the asymptotic limit. Right: The VM is used in the FM case ($J_z=1/2$) in order to investigate the scaling of the half-system entanglement entropy, hinting at the existence of a critical point when extrapolating to the asymptotic limit.}
\label{fig:IsingChainNN}
\end{figure}

\begin{equation}
\pazocal{H}=-J_z\sum\limits_{i=1}^{n-1}  \sigma^{(i)}_z \sigma^{(i+1)}_z - h\sum\limits_{i=1}^n \sigma^{(i)}_x,
\label{eq:IsingChainNN}
\end{equation}
where $J_z >0$ ($J_z<0$) corresponds to FM (AFM) coupling, and $h$ tunes the transverse field strength. For \Cref{eq:IsingChainNN}, the variational method is taken with the effective Hamiltonian $\tilde{\pazocal{H}}:=-(n-1)J_z\sigma_z\otimes\sigma_z-nh(\sigma_x\otimes\mathbbm{1}+\mathbbm{1}\otimes\sigma_x)/2$. Similar to the previous sections, in \Cref{fig:IsingChainNN} we compare the VM with ED for low number of particles. As expected, the VM yields almost orthogonal solutions in the AFM region, while it provides fidelities close to unity in the FM region. Still, slight discrepancies arise in the FM case. In particular, one observes that in the vicinity of what is a critical point (in the asymptotic limit) the fidelity drops, nevertheless still providing a good upper bound to the ground state energy (see \Cref{fig:IsingChainNN}). We remark that, by using the VM to determine the half-system entropy scaling in the FM case ($J_z=1/2$), we observe an anomaly at $h\approx 1$ (see \Cref{fig:IsingChainNN}), which signals the presence of critical point. Therefore, despite the discrepancy in fidelity, the FM critical point can still be well approximated by our VM.

\subsubsection{XXZ model under a transverse field}
\label{sec:XXZchain}
In \Cref{sec:IsingNN} we have looked at a spin system in one dimension and with nearest-neighbour interactions. This model is actually solvable via Jordan-Wigner transformation \cite{JordanWigner1928}, as it can be equivalently described as a system of free fermions (see e.g. \cite{TuraPRX2017}). Here we consider the validity of our variational ansatz beyond the free fermion scope, 
in an XXZ spin chain with an homogeneous magnetic field in the X direction:
\begin{equation}
\pazocal{H}=-J\sum\limits_{\langle i,j\rangle}\left( \sigma_x^{(i)}\sigma_x^{(j)}+\sigma_y^{(i)}\sigma_y^{(j)}+ \Delta\sigma_z^{(i)}\sigma_z^{(j)}\right)+\sum\limits_i h \sigma_x^{(i)} \;,
\label{eq:XXZChain}
\end{equation}
where we take $J=1$ ($J=-1$) for the FM (AFM) couplings, $\Delta $ marks the anisotropy (with $\Delta=1$ corresponding to the isotropic case, the XXX model) and $h$ tunes the transverse field strength. In this case, the effective Hamiltonian for the variational method is $\tilde{\pazocal{H}}:=J\left( (n-1)\left(\sigma_x \otimes \sigma_x +\sigma_y \otimes \sigma_y +\Delta \sigma_z\otimes \sigma_z\right)+nh(\sigma_x\otimes\mathbbm{1}+\mathbbm{1}\otimes\sigma_x )/2 \right)$.

\begin{figure}[h!]
\begin{minipage}{.49\textwidth}
  \centering
  \includegraphics[width=0.8\linewidth]{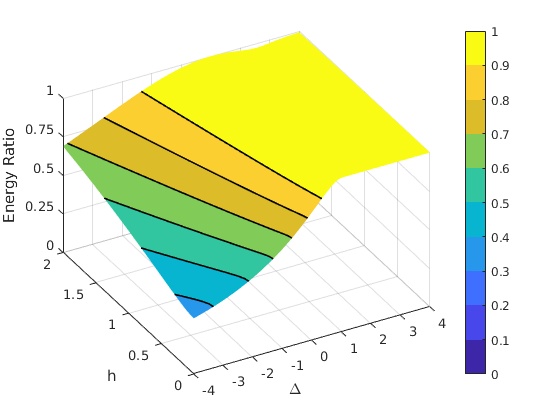}
\end{minipage}
\begin{minipage}{.49\textwidth}
	\centering
	\includegraphics[width=0.8\linewidth]{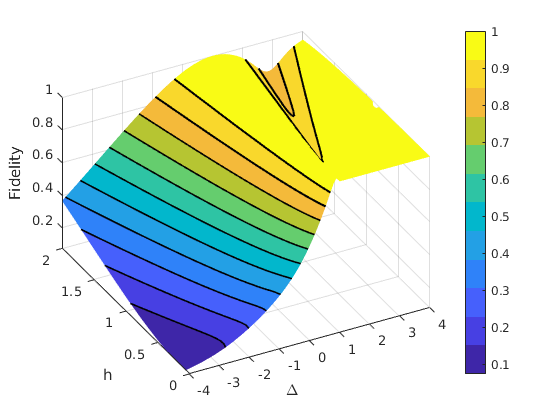}
\end{minipage}
\begin{minipage}{.49\textwidth}
  \centering
  \includegraphics[width=0.8\linewidth]{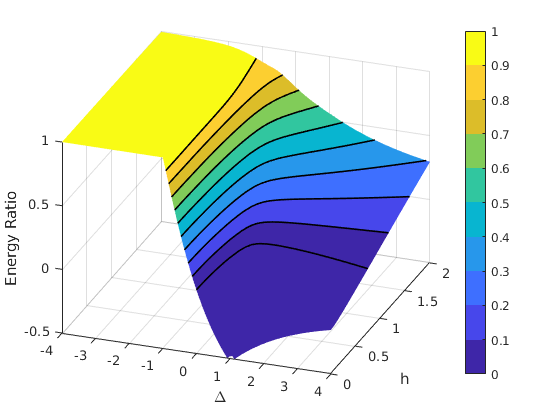}
\end{minipage}
\begin{minipage}{.49\textwidth}
	\centering
	\includegraphics[width=0.8\linewidth]{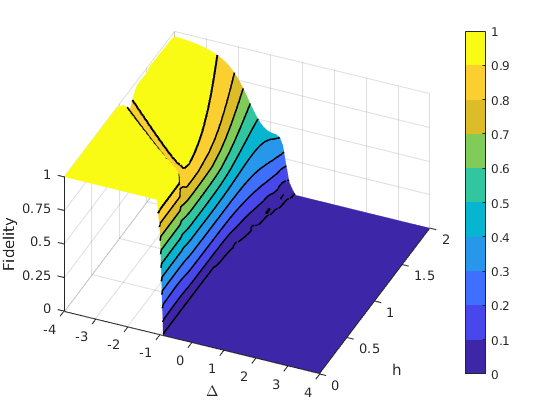}
\end{minipage}
\caption{Numerical results for the XXZ Hamiltonian \eqref{eq:XXZChain} with $n=10$, compared to exact diagonalization. The first row corresponds to FM interactions ($J=1$) while the second row corresponds to AFM interactions ($J=-1$). Left: Ground energy ratio. Right: Ground state fidelity. We observe how the ground state fidelity and energy are well approximated for values $\Delta \gtrsim 1$ ($\Delta \lesssim -1$) when considering FM (AFM) interactions. In such regime it provides close to exact results except in the vicinity of the phase transition between regions $2-3$ and $2-4$ of Figure 1 of \cite{DmitrievJETP2002}.}
\label{fig:XXZChain}
\end{figure}

In \Cref{fig:XXZChain} we compare the ground states obtained from our VM with those obtained from ED, in terms of relative energy and fidelity. For this case we observe that the VM provides a faithful approximation for values $\Delta \gtrsim 1$ ($\Delta \lesssim -1$) when considering FM (AFM) interactions. In particular, in such a regime the VM yields exact results except around a line which likely corresponds to critical points in the asymptotic limit \cite{DmitrievJETP2002, AlcarazJPA1995}.

\subsubsection{Ferromagnetic XXZ with periodic boundary conditions}
\label{sec:ferroXino}
\begin{figure}[h!]
\centering
\begin{minipage}{.32\textwidth}
  \centering
  \includegraphics[width=1\linewidth]{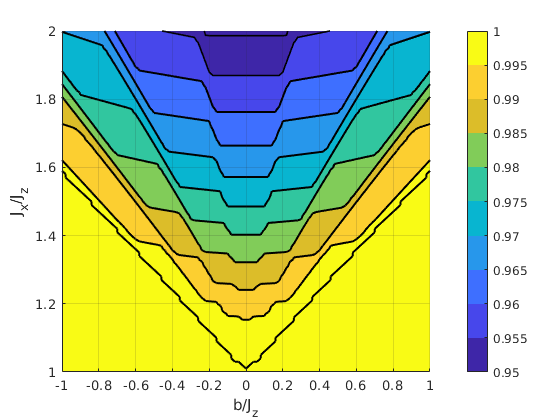}
\end{minipage}
\begin{minipage}{.32\textwidth}
	\centering
	\includegraphics[width=1\linewidth]{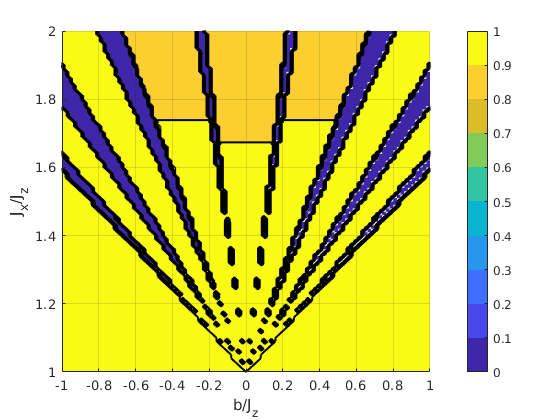}
\end{minipage}
\begin{minipage}{.32\textwidth}
	\centering
	\includegraphics[width=1\linewidth]{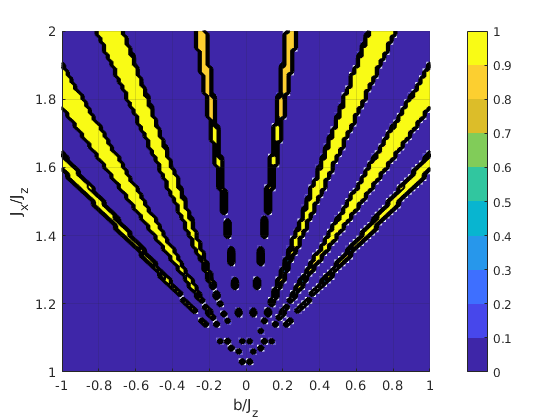}
\end{minipage}
\caption{Numerical results for the XXZ model \Cref{eq:Villa} with $n=10$, compared to exact diagonalization. Left: Energy ratios. Center: Ground state fidelity. Right: First excited state fidelity. We note that the overlap of the right and center figures would cover the whole phase diagram. }
\label{fig:ferroXinoFidelity}
\end{figure}

Let us now consider the following instance of an XXZ model: a periodic anisotropic ferromagnetic spin-1/2 chain, placed in an homogeneous magnetic field in the \textit{z} direction. This model is described by the Hamiltonian
\begin{equation}
\pazocal{H} = -J_{x}\sum\limits_{i=1}^n \left(\sigma_x^{(i)}\otimes \sigma_x^{(i+1)}+\sigma_y^{(i)}\otimes \sigma_y^{(i+1)}\right)-J_z\sum\limits_{i=1}^n \sigma^{(i)}_z\otimes \sigma^{(i+1)}_z+b\sum\limits_{i=1}^n \sigma_z^{(i)},
\label{eq:Villa}
\end{equation}
where $\sigma^{(n+1)}=\sigma^{(1)}$, $J_x, J_z \geq 0$ are the exchange coupling constants, and $b$ tunes the strength of the external magnetic field. In Ref.~\cite{ZhouPRA2011} it was investigated the fidelity of preparation of Dicke states as ground states of \Cref{eq:Villa}. In \Cref{table:ferroXino} we show the ground state distribution predicted with the perturbative results of \cite{ZhouPRA2011}. Our variational ansatz comes as a natural tool to benchmark the fidelity of the preparation of Dicke states as the ground states of the Hamiltonian in \Cref{eq:Villa}. In \Cref{fig:ferroXinoFidelity} we see that, up to $n=10$, the fidelity remains $>85\%$, which is consistent with the predictions of \cite{ZhouPRA2011}. For this case, in the VM we take the effective Hamiltonian $\tilde{\pazocal{H}}:=-n\left(J_x \left(\sigma_x\otimes \sigma_x+\sigma_y\otimes \sigma_y\right)+J_z \sigma_z\otimes \sigma_z\right)+n(\sigma_z\otimes \mathbbm{1}+\mathbbm{1}\otimes \sigma_z)/2$.

\begin{table}[h!]
\centering
\begin{tabular}{|c|c|}
\hline
$b$ & Ground state \\
\hline
$b <-\Delta J$ & $\ket{D_0^n}$ \\
$-\frac{n-2k+1}{n-1}\Delta J<b<-\frac{n-2k-1}{n-1}\Delta J$ & $\ket{D_k^n}$, $0<k<n$\\
$\Delta J < b$ & $\ket{D_n^n}$ \\
\hline
\end{tabular}
\caption{Ground state for the model in \Cref{eq:Villa}, as a function of the magnetic field $b$ and the coupling parameter $\Delta J=J_x-J_z$, according to the perturbative results presented in \cite{ZhouPRA2011}. In \Cref{fig:ferroXinoFidelity} we recover and strengthen the result.}
\label{table:ferroXino}
\end{table}

The perturbative prediction from \cite{ZhouPRA2011} splits the phase diagram among different regions, each having substantial overlap to a different Dicke state. However, there are already some discrepancies observed: In between these regions, the approximation with Dicke states does not have good overlap with the ground state (see Figure 1 in \cite{ZhouPRA2011}), however it has good overlap with other basis Dicke states (see Figure 2 in \cite{ZhouPRA2011}). Here, our method enables us to understand this discrepancy from a different perspective: in \Cref{fig:ferroXinoFidelity} we see that the regions in which the perturbative approach fails actually correspond to a good overlap for the first excited state and a Dicke state. The low-end of the spectrum of the Hamiltonian considered has a good overlap with the symmetric space. However, it can happen that the VM chooses to approximate the first excited state instead of the ground state if the energy obtained becomes more favorable. This depends on both the overlap with the ground space and the energy gap of the Hamiltonian. Let us denote $E_0$ and $E_1$ the ground and first excited state energies, respectively. Let us also denote $F_0$ ($F_1$) the fidelity between the ground state (first excited state) and the symmetric space. The discontinuities may happen when $F_0 E_0 = F_1 E_1$. Indeed, in \Cref{fig:ferroXinoFidelity} we observe that, while the energy ratio is smooth, the fidelity may suddenly drop to zero or jump to almost one due to the above mentioned reason.

For the larger $n$ limit, the ground state of \Cref{eq:Villa} can be found exactly using matrix product states (MPS) and the density matrix renormalization group (DMRG) algorithm. In order to compute the fidelity of our variational solution with respect to the exact one, the most straightforward way is to contract the MPS representation of the ground state with the solution of the variational method. However, the latter is not given in a MPS form and it therefore has to be converted in a MPS form. In order to do so, we need to establish a correspondence between arbitrary superpositions of Dicke states and MPS. To the best of our knowledge, in general such a correspondence has not been established before, and this result may be of independent interest. Therefore, we have devoted a full section (\Cref{sec:MPS}) to it.

\subsubsection{Many-body spin-$1$ Hamiltonian with collective interactions}
\label{sec:Kus}

Finally, let us illustrate that our proposed variational method can be also easily applied to systems with local Hilbert space of dimension $d>2$. To do so, we consider the three-orbital Lipkin-Meshkov-Glick Hamiltonian \cite{MeredithPRA1988} (equivalently, the generalisation of the Lipkin Hamiltonian as proposed in \cite{GnutzmannJPhysA1999}). Similarly to \Cref{subsec:LMG}, the variational ansatz is expected to recover exact results due to the long-range interactions resulting in ground states with permutation symmetry. The model is constructed by $n$ identical but distinguishable three-level atoms, and it is also commonly used in nuclear shell models. It can also arise for three-level atoms collectively coupled to electromagnetic field modes of a cavity. Concretely, the model is described by the Hamiltonian:

\begin{equation}
\pazocal{H}=a \left(\pazocal{S}_{00}-\pazocal{S}_{22}\right)+b\sum\limits_{i\neq j}\pazocal{S}_{ij}^2,
\label{eq:LipkinKus}
\end{equation}
where $\pazocal{S}_{ij}=\sum\limits_{l=1}^n=\tau_{ij}^{(l)}$ with $\tau_{ij}=\ket{i}\!\bra{j}$ for $i,j=\{0,1,2\}$. In this case the effective Hamiltonian used in the SDP is $\tilde{\pazocal{H}}:=na\left( \pazocal{S}_{00}\otimes\mathbbm{1}+\mathbbm{1}\otimes\pazocal{S}_{00}-\pazocal{S}_{22}\otimes\mathbbm{1}-\mathbbm{1}\otimes\pazocal{S}_{22}\right)/2+{n\choose 2}b\sum\limits_{i\neq j}\left(\pazocal{S}_{ij}\otimes\pazocal{S}_{ij}+(\pazocal{S}_{ij}^2\otimes\mathbbm{1}+\mathbbm{1}\otimes\pazocal{S}_{ij}^2)/2\right)$.

\begin{figure}[h!]
\centering
\begin{minipage}{.33\textwidth}
  \centering
  \includegraphics[width=1\linewidth]{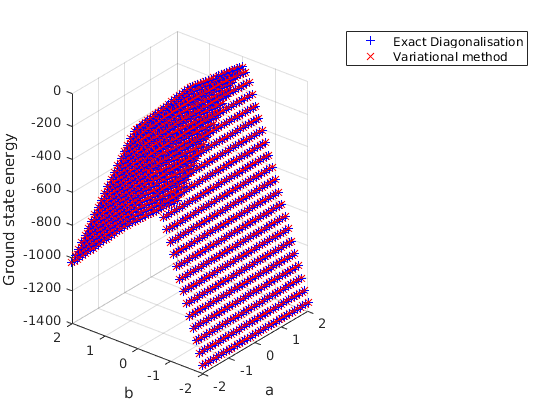}
\end{minipage}
\begin{minipage}{.33\textwidth}
  \centering
  \includegraphics[width=1\linewidth]{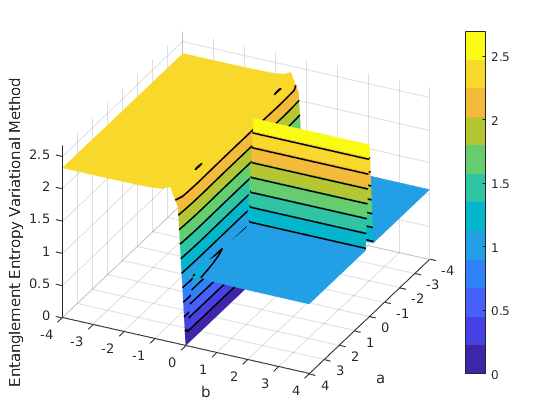}
\end{minipage}
\begin{minipage}{.32\textwidth}
  \centering
  \includegraphics[width=1\linewidth]{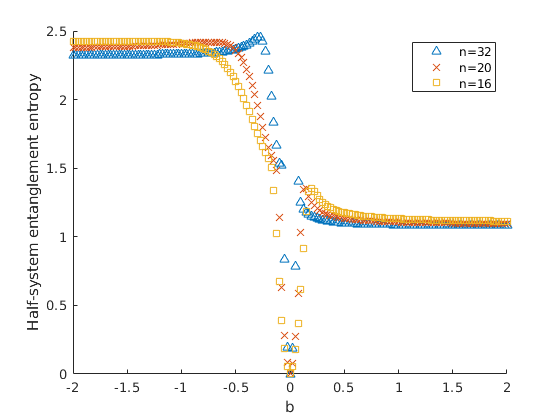}
\end{minipage}
\caption{Numerical results for the three-level generalized Lipkin Hamiltonian \Cref{eq:LipkinKus} with $n=32$. Left: The VM provides exact ground state up to numerical error of the solver ($\approx 10^{-14}$ using SeDuMi \cite{SeDuMi}). Center: Half-system entanglement entropy obtained through the compatibility conditions \Cref{eq:Cocu}, capturing features of phase transitions. Right: Half-system entanglement entropies for different number of particles. The scalability can be used to extrapolate the peak anomalies to the asymptotic limit.
}
\label{fig:LipkinKus}
\end{figure}

In \Cref{fig:LipkinKus} we show that the variational ansatz reproduces exactly the ground state energy, as expected. Furthermore, we use the compatibility conditions to obtain the half-system entanglement entropy, which is useful to provide insights about the phase diagram of the model.

\subsubsection{Benchmarking performance with existing methods}
\label{sec:time}

One of the most appealing features of the variational method proposed here is its ability to yield results for large system sizes with very modest time and memory requirements (and, consequently, energy consumption). Indeed, since the computations take place in the symmetric space, its dimension grows only polynomially with the system size. In particular, it is linear for qubits, quadratic for qutrits, etc. In the previous sections we have argued that the method yields results that capture traces of some quantities of physical interest. Therefore, it can be a good candidate to a first order exploration before trying more numerically-intensive results. To make this comparison quantitative, we here benchmark the computational requirements of the method with other existing techniques.

In this section we briefly comment on the time, memory and energy consumption devoted to the variational ansatz. The runtime of the variational method can be split in two steps: 1) to precompute the $A$ matrices in \Cref{eq:AlbertJorquera} for a fixed $n$, $m$ and $d$ and 2) to load and solve the SDP. The most expensive task, both in time and memory, is to compute the compatibility constraints. Hence, in order to agilitate the process, one would first preallocate and store the compatibility constraints for a fixed number of particles $n$, of local Hilbert space dimension $d$ and with RDMs of size $m$. Then, once the compatibility constraints are preallocated, one can scan the phase diagram of the desired parametrized Hamiltonian model just by loading the compatibility constraints and proceeding to solve the corresponding SDP.

\begin{figure}[h!]
  \centering
  \begin{minipage}{.32\textwidth}
  \centering
  \includegraphics[width=1\linewidth]{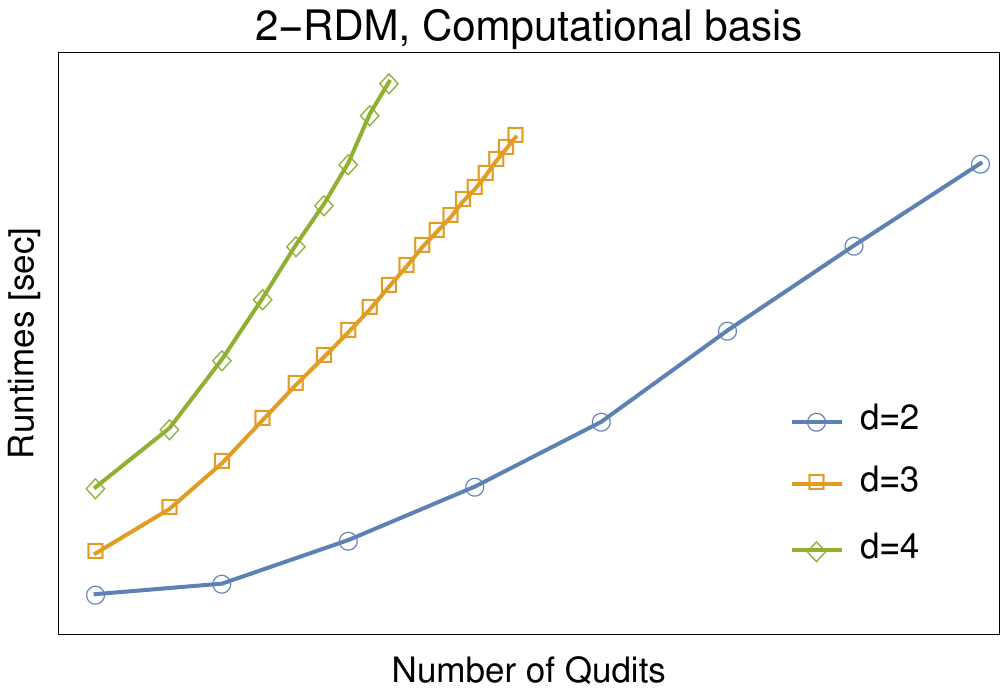}
  \end{minipage}
  \begin{minipage}{.32\textwidth}
  \centering
  \includegraphics[width=1\linewidth]{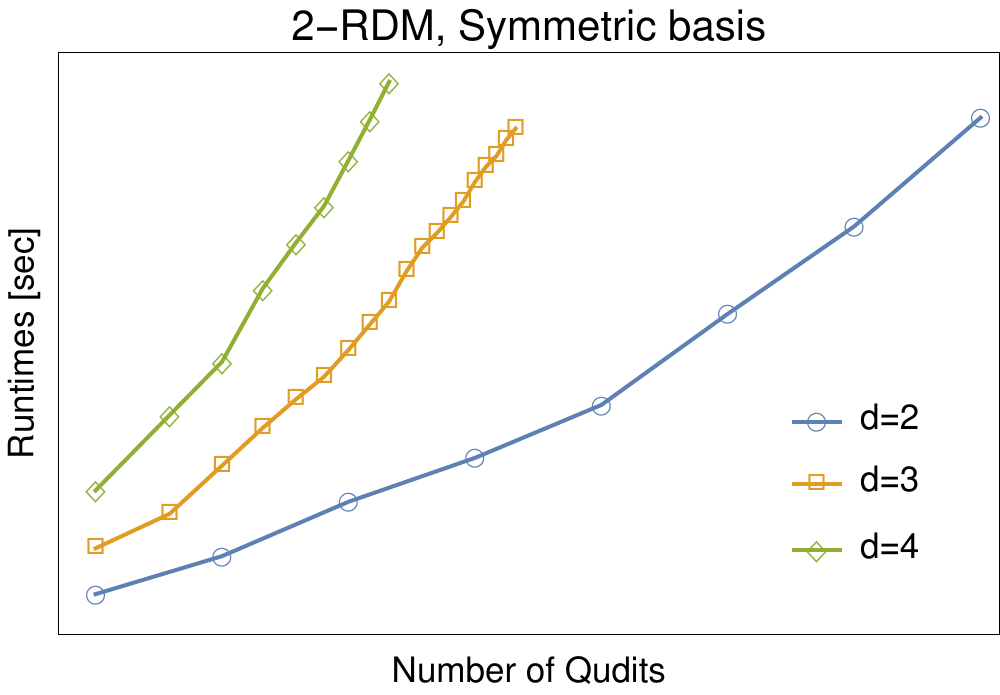}
  \end{minipage}  
  \begin{minipage}{.32\textwidth}
  \centering
  \includegraphics[width=1\linewidth]{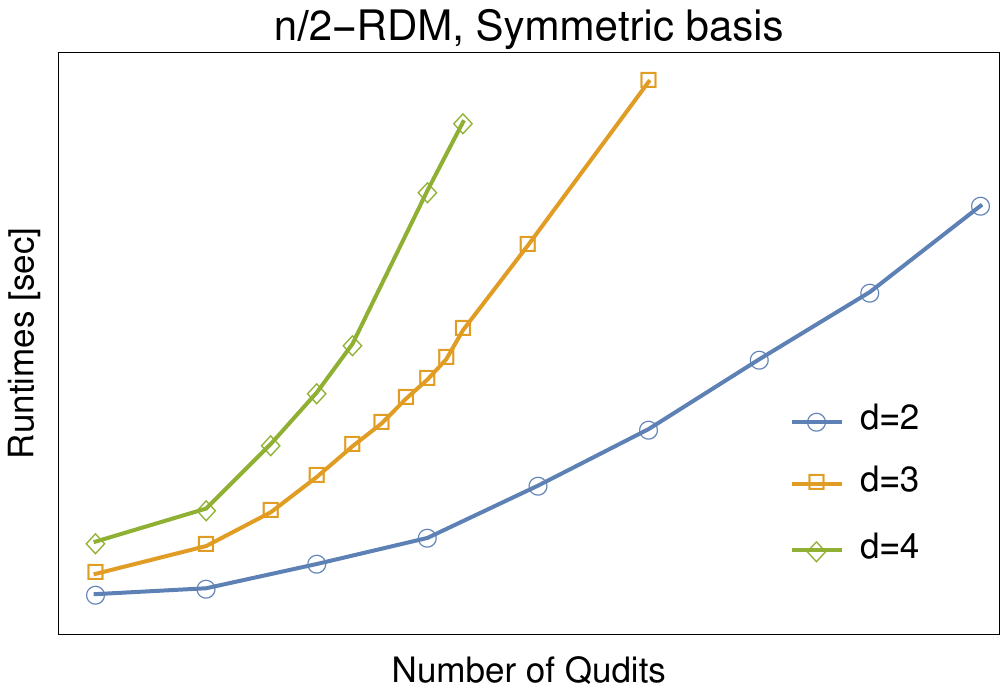}
  \end{minipage}
  \caption{Runtimes observed in order to prellocate the $2$-RDMs and $n/2$-RDMs compatibility constraints in the computational and symmetric basis. Apart from the memory storage advantage, it is observed that the symmetric spaces offers a significant advantage also in runtimes. The computations have been carried out on a 64-bit operating system with 32GB ram and a 3.70GHz processor. No parallelization has been used, which can be easily be implemented, significantly speeding up the process. The runtimes might slightly vary at each run and are not meant to be taken as exact, but as an illustration of their order.}
  \label{fig:RuntimesConstraints}
\end{figure}

In \Cref{fig:RuntimesConstraints} we present a representative sample of the computing runtimes we have observed in order to prellocate the compatibility constraints of the $2$-RDMs and half-system $n/2$-RDMs for different number of particles $n$ and different local Hilbert space dimensions $d$. We have considered the $2$-RDMs compatibility constraints case both in the computational basis (\Cref{eq:Deco}) and the symmetric basis (\Cref{eq:Cocu}). Apart from requiring less memory, one observes a clear advantage in runtime when obtaining the compatibility constraints directly in the symmetric basis, as expected. Therefore, for the variational method it is desirable to project the effective Hamiltonian onto the symmetric subspace. For the $n/2$-RDMs case, we have considered only the symmetric representation which decreases the memory storage limitations in order to find, for instance, the half-system entanglement entropies. An additional comment is in order: For a constant value of $d$, note that the multinomial coefficients in \Cref{eq:Deco} or \Cref{eq:Cocu} do not require a full expansion of the factorials, but there exist closed analytical formulas for them (see e.g. \cite{AnnPhys}). This has been taken into consideration in our calculations. Furthermore, it is desirable to apply such closed expressions, not only for speed, but more importantly for numerical stability issues (quotients of factorials of large numbers may give problems in floating-point arithmetic if these numbers are of the order of $\sim 100$).

\begin{figure}[h!]
  \centering
  \begin{minipage}{.32\textwidth}
  \centering
  \includegraphics[width=1\linewidth]{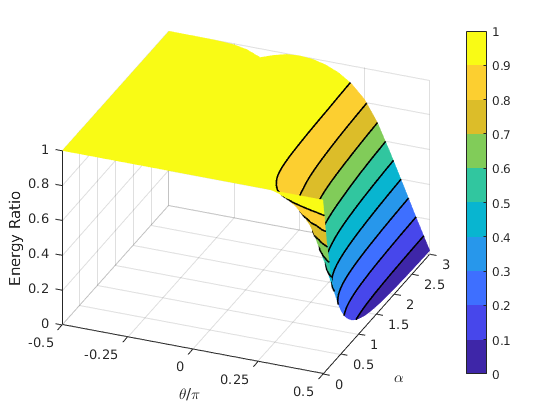}
  \end{minipage}
  \begin{minipage}{.32\textwidth}
  \centering
  \includegraphics[width=1\linewidth]{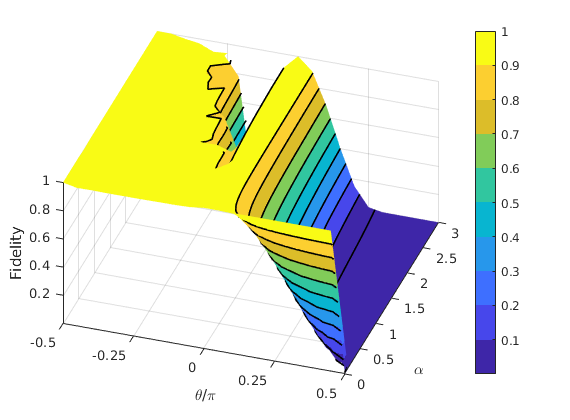}
  \end{minipage}
  \begin{minipage}{.32\textwidth}
  \centering
  \includegraphics[width=1\linewidth]{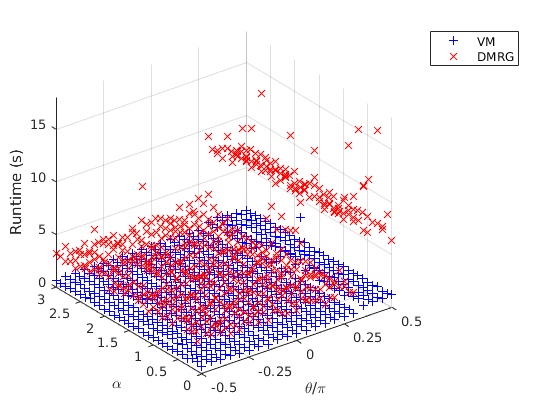}
  \end{minipage}  
  \caption{Numerical results of the VM for the Ising Hamiltonian \Cref{eq:IsingChainPowerLaw} with $n=64$, compared with DMRG. The DMRG algorithm follows \cite{PigaPRL2019,CrosswhitePRB2008,FroewisPRA2010} increasing the bond dimension up to $20$. Left and Center: Energy ratio and fidelity of the VM solution with respect to the DMRG solution. In order to compute the fidelity we use the result in \Cref{sec:MPS} to transform the symmetric basis representation of the VM solution into a matrix product state representation. Right: Runtimes comparison to achieve convergence with DMRG and VM method. The VM is significantly faster, making it a good candidate for a first rough exploration of large phase diagrams, and to upper bound ground state energies. Note that runtimes might slightly vary at each run. The total runtimes are $525.04$s ($1$s preallocation) for the VM and $2854.25$s for the DMRG. With a power consumption of $425$W in our workstation, amounts to an environmental impact of around $18$g vs $100$g of CO${}_2$ into the atmosphere for the VM vs the DMRG methods (we have taken the $0.296$ EU coefficient of kWh to kg of CO${}_2$ given by the European Environment Agency for $2016$). We remark that the DMRG is an extremely optimized and efficient method that cannot be applied beyond $1$ geometric dimension. In these cases, the benchmark with existing methods would be separated by even more orders of magnitude.}
  \label{fig:RuntimesSDP}
\end{figure}

In \Cref{fig:RuntimesSDP} we present some of the computing runtimes in order to load the constraints and solve the SDP for the Ising chain with decaying power-law interactions previously considered in \Cref{sec:IsingPowerLaw}. We have considered the constraints and effective Hamiltonian in the symmetric basis, and in order to solve the SDP we have set the solver SDPT3 \cite{SDPT3} to its maximal precision providing a numerical error up to $\pazocal{O} (10^{-14})$ when the variational ansatz can reach the exact solution. We have carried out the comparison with the solution provided by DMRG. In order to find the fidelity between the DMRG solution and the VM solution, we have used the auxiliary results developed in \Cref{sec:MPS} in order to represent the VM solution as a translationally invariant diagonal matrix product state.

\subsection{Bell non-local correlations}
\label{sec:Nonlocal}
The proposed variational ansatz is also convenient to investigate Bell non-local correlations in many-body systems.
The permutational symmetry naturally synergizes with the so-called 2-body Permutation Invariant Bell Inequalities (PIBIs) presented in \cite{SciencePaper}. These type of Bell inequalities involve at most 2-body correlation functions, and some of them are violated by symmetric states \cite{AnnPhys}. One can now consider two approaches: On the one hand, to obtain the quantum state that gives the maximal violation of such equalities within the variational ansatz. On the other hand, to find quantum states that also have Bell correlations by using the VM to appoximate the ground state of a many-body Hamiltonian. In the latter case, the Hamiltonian considered need not correspond to the Bell operator \cite{TuraPRX2017}. It is worth mentioning that the measurement settings might need to be optimized in order to increase the visibility of the Bell correlations.

\subsubsection{Optimizing permutationally invariant two-body Bell inequalities}
We first focus on two particular classes of 2-body PIBI. These inequalities satisfy the following condition for all correlations that can be described under local-realism assumptions (meaning that their violation signals the presence of non-local correlations):

\begin{equation}
-2\pazocal{S}_0+\frac{1}{2}\pazocal{S}_{00}-\pazocal{S}_{01}+\frac{1}{2}\pazocal{S}_{11}\geq -2n
\label{eq:ineq6}
\end{equation}
and
\begin{equation}
(n \bmod 2)(n-1)(n\pazocal{S}_0+\pazocal{S}_1)+{n\choose 2}\pazocal{S}_{00}+n\pazocal{S}_{01}-\pazocal{S}_{11}\geq {n \choose 2}(n + 2 + n \bmod 2),
\label{eq:ineqDicke}
\end{equation}
where $\pazocal{S}_k=\sum_{i=1}^n \braket{\pazocal{M}^{(i)}_k}, \pazocal{S}_{kl}=\sum_{i\neq j} \braket{\pazocal{M}_k^{(i)}\pazocal{M}_l^{(j)}}$ are the one- and two-body symmetric correlators with $\pazocal{M}_k^{(i)}$ denoting the measurement in direction indexed by $k=\{0,1\}$ corresponding to particle $i$. The first inequality \Cref{eq:ineq6} is particularly fitted to detect non-local correlations in superpositions of Dicke states, while the second inequality \Cref{eq:ineqDicke} is tailored to detect non-local correlations in half-filled pure Dicke states \cite{AnnPhys, FadelPRL2017}.

In order to know if there exists a quantum state that violates such Bell inequalities, one still needs to find appropiate $n$ pairs of measurement settings. This gives rise to the so-called Bell operator, a quantum observable in the $n$-partite Hilbert space whose expectation value with respect to a quantum state corresponds to the value of the Bell inequality for the chosen measurement settings. In general, finding such measurements consists in a very demanding non-convex optimization problem with no trivial solution. However, since the variational ansatz provides a global state, the complexity of the problem gets greatly reduced when we restrict the optimization to the case where all the measurement settings are the same for each party, in the same reference frame. This gives rise to a permutationally invariant Bell operator, whose extremal values within the symmetric space we be found using our variational method.

Contrary to previous approaches (see e.g. \cite{AnnPhys, PhDTura}), where one could use representation theory methods such as Schur-Weyl duality to project the permutationally invariant Bell operator onto the different symmetric blocks (see also \cite{TavakoliPRL2019}), here the VM circumvents this intermediate step: It is enough to consider the effective Hamiltonian
\begin{equation}
-2n\rm{Tr}\Big(\left(\pazocal{M}_0\otimes\mathbb{I}+\mathbb{I}\otimes\pazocal{M}_0\right)\sigma\Big)+{n\choose 2}\rm{Tr}\Big(\left(\pazocal{M}_0\otimes\pazocal{M}_0-2\pazocal{M}_0\otimes\pazocal{M}_1+\pazocal{M}_1\otimes\pazocal{M}_1\right)\sigma\Big),
\label{eq:2RDMopt}
\end{equation} 
where $\sigma$ can be the 2-RDM obtained with the variational ansatz. We parametrize the measurements as $\pazocal{M}_k:=\sin \left(\theta_k\right)\sigma_x+\cos\left(\theta_k\right)\sigma_z$, where $k\in\{0,1\}$ and $\sigma_x, \sigma_z$ are the Pauli matrices, and use the VM to find the symmetric state minimizing the energy of the effective Hamiltonian. We note that this approach becomes particularly useful in the case of large $d$, since the number of blocks arising from the symmetry-adapted basis
increases with $d$. We also remark that, since one can always apply a dual $U^{\otimes n}$ symmetry to both state and measurements without departing from the symmetric space, it is enough to optimize over the difference between measurement directions, \textit{e.g.} $\theta_0-\theta_1$. Furthermore, since \Cref{eq:ineq6} and \Cref{eq:ineqDicke} have two inputs and two outputs per party, Jordan's lemma guarantees that using $d=2$ is sufficient to find find its maximal violation.

\subsubsection{Looking for Bell correlations in a direction specified by a Hamiltonian}

\begin{figure}[h!]
\centering
\begin{minipage}{.32\textwidth}
  \centering
  \includegraphics[width=1\linewidth]{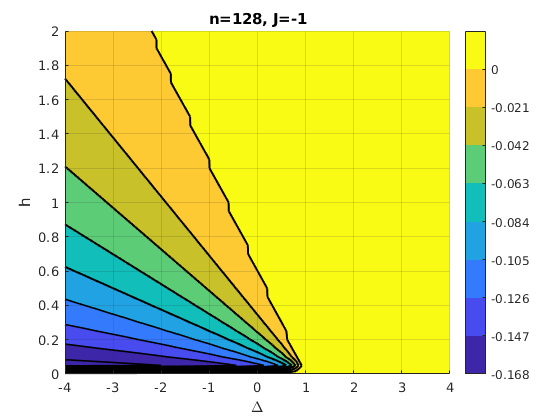}
\end{minipage}
\begin{minipage}{.32\textwidth}
  \centering
  \includegraphics[width=1\linewidth]{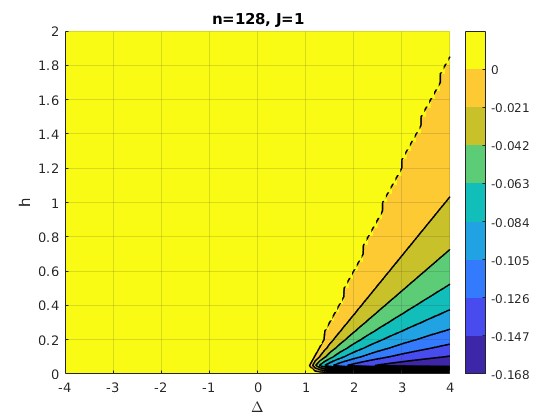}
\end{minipage}
\begin{minipage}{.32\textwidth}
  \centering
  \includegraphics[width=1\linewidth]{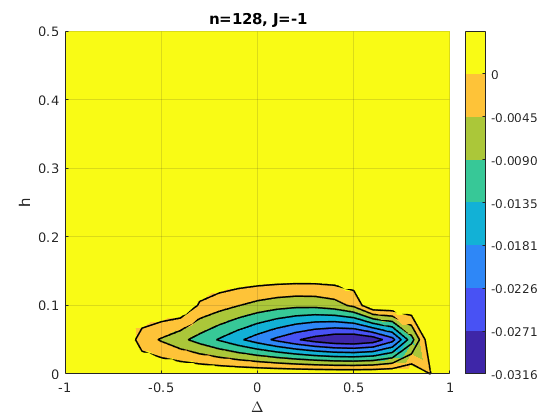}
\end{minipage}
\caption{Detection of non-local correlations using inequality \Cref{eq:ineq6} in the XXZ chain \Cref{eq:XXZChain} with $n=128$ particles. Left: Bell inequality value normalized to the classical bound using the VM with \Cref{eq:2RDMopt} for $J=-1$ FM couplings. Values below zero indicate nonlocal correlations in the VM solution. Center: Same for $J=1$ AFM couplings. Right: Zoom-in of the region with largest relative violation for the FM case with largest fidelities according to \Cref{fig:XXZChain}, conditioned on finding Bell correlations with the VM. Here the expectation values have been computed using the actual ground state given by the DMRG solution with the (non-symmetric) Bell operator \Cref{eq:2RDMoptfree}. We have chosen a bond dimension of $32$ for the DMRG solution.
}
\label{fig:XXZ_NL}
\end{figure}

Here we propose a two-step process to find Bell correlations in the ground state of Hamiltonians of physical interest (e.g. an XXZ chain). First, we use the VM to do a quick scan over the parameter space of a given Hamiltonian family, in order to find potential candidates whose ground state might display Bell correlations, see \Cref{fig:XXZ_NL}. If we found Bell correlations with the VM, we would have obtained a symmetric state displying them, albeit we have no guarantee of the fidelity with the ground state of the model at this point.
Then, we narrow down the search by computing the actual ground state with other methods, such as DMRG, also in \Cref{fig:XXZ_NL}. As we show in \Cref{fig:XXZ_NL}, we observe for the first time, to the best of our knowledge, that the ground state of such an XXZ chain violates the Bell inequality \Cref{eq:ineq6}, thus exhibiting Bell correlations, in the corresponding parameter regime.
Bear in mind that in order to carry out the measurement optimization one can no longer consider a single 2-RDM as we posed in \Cref{eq:2RDMopt}, but one has to sum over all the different 2-RDMs $\sigma_{ij}$ obtained with the DMRG:
\begin{equation}
-2\sum_{i=1}^n\rm{Tr}\Big(\left(\pazocal{M}_0^{(i)}\otimes\mathbb{I}+\mathbb{I}\otimes\pazocal{M}_0^{(i)}\right)\sigma_i\Big)+\sum_{i<j}\rm{Tr}\Big(\left(\pazocal{M}_0^{(i)}\otimes\pazocal{M}_0^{(j)}-2\pazocal{M}_0^{(i)}\otimes\pazocal{M}_1^{(j)}+\pazocal{M}_1^{(i)}\otimes\pazocal{M}_1^{(j)}\right)\sigma_{ij}\Big) \;.
\label{eq:2RDMoptfree}
\end{equation}
Note that, since the RDMs are fixed in this case (by the exact solution), the measurement directions should not be restricted to the XZ plane in the Bloch sphere, but allowed to point in any direction. Furthermore, the measurement settings for each party do not need to coincide.

\begin{figure}[h!]
\centering
\begin{minipage}{.49\textwidth}
  \centering
  \includegraphics[width=0.8\linewidth]{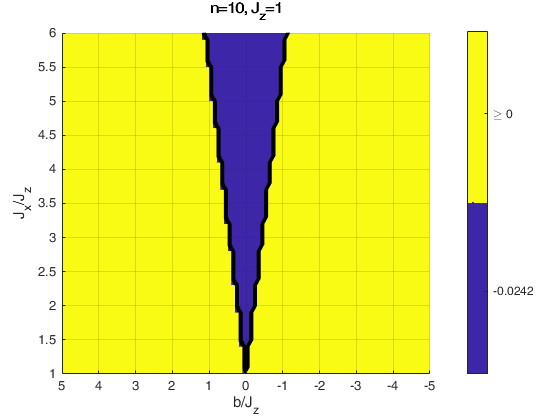}
\end{minipage}
\begin{minipage}{.49\textwidth}
  \centering
  \includegraphics[width=0.8\linewidth]{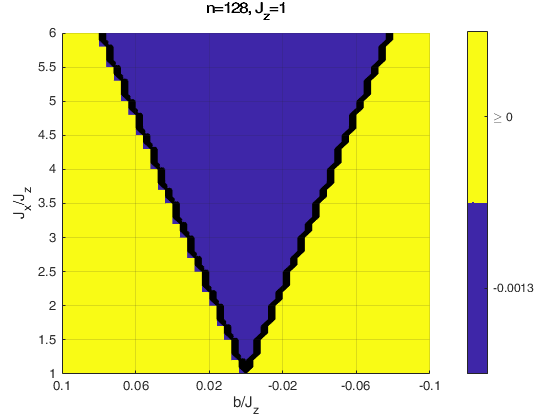}
\end{minipage}
\caption{Detection of non-local correlations of the solution given by the VM using inequality \Cref{eq:ineqDicke} for the ferromagnetic XXZ chain with periodic boundary conditions and under a longitudinal magnetic field $b$ \Cref{eq:Villa} with $n=10$ and $n=128$ particles respectively. We observe non-local correlations not only in the region where the variational ansatz approximates the ground state with pure half-filled Dicke states, but also where such ansatz approximates the first excited state. Negative numbers indicate the presence of Bell correlations.}
\label{fig:XXZxino_NL}
\end{figure}

Similarly, we have seen in \Cref{sec:ferroXino} that pure Dicke states provide a good approximation of the ground state for such ferromagnetic XXZ chain with periodic boundary conditions and longitudinal magnetic field $b$. As we have previously mentioned, inequality \Cref{eq:ineqDicke} is tailored to half-filled Dicke states which happen to approximate the ground state in the range $-\frac{1}{n-1}(J_x - J_z)< b <\frac{1}{n-1}(J_x - J_z)$. Therefore, in such region we expect to witness non-local correlations with \Cref{eq:ineqDicke}. Indeed, in \Cref{fig:XXZxino_NL} we show the witnessed nonlocality where we have used the variational ansatz to approximate the ground state and optimize using \Cref{eq:2RDMopt}. We also observe that nonlocality detection goes beyond the specified region where the half-filled Dicke state approximates the ground state. As previously discussed in \Cref{sec:ferroXino}, such extra range of nonlocality detection seems to arise from the variational ansatz approximating the first excited state instead of the ground state.%
	
\subsection{Generically expressing symmetric states as TI diagonal MPS}
\label{sec:MPS}
In this section we present an analytical method to generically represent any $n$-qubit Dicke state with a translationally-invariant (TI) matrix product state (MPS).
More precisely, the goal is that, given a state of the form
\begin{equation}
 \ket{\psi} = \sum_{k=0}^{n} d_k \ket{D_n^k},\qquad d_k \in \mathbbm{C},
 \label{eq:AnsuFati}
\end{equation}
find two matrices $A_{0}, A_1 \in {\pazocal M}_{D\times D}(\mathbbm{C})$ such that
\begin{equation}
 \ket{\psi} = \sum_{(i_1, \ldots, i_n) \in \{0,1\}^n} \psi_{(i_1, \ldots, i_n)} \ket{i_1, \ldots, i_n} = \sum_{(i_1, \ldots, i_n) \in \{0,1\}^n} \mathrm{Tr}[A_{i_1}\cdots A_{i_n}] \ket{i_1, \ldots, i_n}.
 \label{eq:umtiti}
\end{equation}

Some representations of important symmetric states have been known since their inception. For instance, the GHZ state (up to normalization) can be generated with $D=2$ \cite{OrusAoP2014} using the following TI MPS:
\begin{equation}
 A_0=\left(
 \begin{array}{cc}
  1&0\\0&0
 \end{array}
 \right), \qquad
 A_1=\left(
 \begin{array}{cc}
  0&0\\0&1
 \end{array}
 \right).
 \label{eq:JoseMariVaquero}
\end{equation}

On the other hand, for the $\ket{W}$ state, there exists no TI representation with bond dimension 2 \cite{Perez-GarciaQIC2007}. However, it does admit (up to normalization) the following non-TI representation \cite{Perez-GarciaQIC2007}:
\begin{equation}
 \left\{
 \begin{array}{ccccc}
 \{A_0^{[i]}, A_1^{[i]}\} &=& \{\sigma^+, \mathbbm{1}\}&\mbox{if}& i < n\\
 \{A_0^{[i]}, A_1^{[i]}\} &=& \{\sigma^+\sigma_x, \sigma_x\}&\mbox{if}& i = n.
 \end{array}
 \right.
 \label{eq:PituAbelardo}
\end{equation}

We observe that in \Cref{eq:PituAbelardo}, all the coefficients of the MPS are either $1$ or $0$. Indeed, some MPS can be used to represent Boolean formula solutions \cite{Biamonte2017}. More generally, the representability of quantum states with MPS of a particular form has deep connections with modern algebraic geometry \cite{SanzPRA2009, SanzJPA2017, SanzSciRep2016}. For instance, the $\ket{W}$ state can be arbitrarily well-approximated with a diagonal TI MPS of bond dimension $D=2$:
\begin{lem}
\label{lem:ElsBocatesDePernilCampofrioALestadiJohannCruyffJaVanA8EurusEsGravissim}
 Let $\varepsilon > 0$ and
 \begin{equation}
 A_0=\left(
 \begin{array}{cc}
  x_0&0\\0&x_1
 \end{array}
 \right), \qquad
 A_1=\left(
 \begin{array}{cc}
  y_0&0\\0&y_1
 \end{array}
 \right),
 \label{eq:GuillermoAmor}
\end{equation} where
\begin{equation}
 \left\{
 \begin{array}{ccc}
 x_0 &=&2^{-1/n}\varepsilon^{-1/[n(n-1)]}\\
 x_1 &=&e^{\mathbbm{i} \pi / n}x_0\\
 y_0&=& 2^{-1/n} \varepsilon^{1/n}\\
 y_1 &=& -e^{\mathbbm{i}\pi/n}y_0.
 \end{array}
 \right.
 \label{eq:vanHaal}
\end{equation}
The $n$-qubit $\ket{W}$ state can be obtained as the limit of $\varepsilon \rightarrow 0$ of the TI MPS given by \Cref{eq:GuillermoAmor}.
\end{lem}
\paragraph{Proof:}

One simply notes that, since the MPS is diagonal, the element corresponding to the physical index $(i_1, \ldots, i_n)$ with $k:=\sum_{j}i_j$ is given by
\begin{equation}
 \psi_{(i_1,\ldots, i_n)}=x_0^{n-k}y_0^k + x_1^{n-k}y_1^{k},
\end{equation}
which amounts to $0$ if $k\equiv 0 \mod 2$ and $\varepsilon^{(k-1)/(n-1)}$ if $k \equiv 1 \mod 2$. Hence, noting
\begin{equation}
 \lim_{\varepsilon \rightarrow 0} \varepsilon^{\frac{k-1}{n-1}} = \left\{
 \begin{array}{ccc}
  1&\mbox{if}&k=1\\
  0&\mbox{if}&k>1
 \end{array}
 \right.
\end{equation}
yields the result. \QED

Inspired by \Cref{lem:ElsBocatesDePernilCampofrioALestadiJohannCruyffJaVanA8EurusEsGravissim}, we propose now a TI MPS of bond dimension $n$ to approximate \emph{generically} any superposition of Dicke states of the form \Cref{eq:AnsuFati}. We propose the following parameterization of $A_{0}$ and $A_{1}$:
\begin{equation}
 A_0 \propto \mathbbm{1}_D; \qquad A_1 = \mathrm{diag}(x_1, \ldots, x_D).
\end{equation}
For simplicity let us denote $A_0 = y\mathbbm{1}$ and $k=\sum_{j}i_j$. It is then easy to see that \Cref{eq:umtiti} leads to the following system of equations:
\begin{equation}
 \sum_{a=1}^D y^{n-k}x_{a}^{k} = \psi_{(i_1, \ldots, i_n)}, \qquad \{i_1, \ldots, i_n\} \in \{0,1\}^n.
\end{equation}
Note that for states of the form of \Cref{eq:AnsuFati} we only require $n+1$ equations. Hence, it is natural to choose $D=n$. This motivates the following Lemma:

\begin{lem}
 Consider the system of Equations
 \begin{equation}
  \left\{
  \begin{array}{ccc}
   x_1+\cdots + x_n &=& z_1\\
   x_1^2+\cdots+x_n^2&=& z_2\\
   \cdots\\
   x_1^n+\cdots+x_n^n&=&z_n
  \end{array}
  \right.,
  \label{eq:Guardiolo}
 \end{equation}
 where $z_1, \ldots, z_n \in \mathbbm{C}$. The solutions of \Cref{eq:Guardiolo} are the roots of the polynomial $P(X)$ defined in \Cref{eq:Pedrito}.
 \label{lem:FrastfurtA3Euros}
\end{lem}

\begin{lem} 
The system of equations that determines the coefficients $y$ and $x_1\ldots x_n$ of the diagonal tensors $A_0$ and $A_1$, used to represent the linear combination of Dicke states \Cref{eq:AnsuFati}, is given by
  \begin{equation}
  \left\{
  \begin{array}{ccc}
   y&=&\sqrt[n]{d_0/n}\\
   x_1+\cdots + x_n &=& \frac{d_1}{y^{n-1} \sqrt{n}}\\
   x_1^2+\cdots+x_n^2&=& \frac{d_2}{y^{n-2} \sqrt{n \choose 2}}\\
   \cdots\\
   x_1^k+\cdots+x_n^k&=& \frac{d_k}{y^{n-k} \sqrt{n \choose k}}\\
   \cdots\\
   x_1^n+\cdots+x_n^n&=&d_n
  \end{array}
  \right..
  \label{eq:LaportaPresidentCatalunyaIndependent}
 \end{equation}
 The value of $y$ is readily determined from the first equation, and the $x$'s are found by finding the roots of the polynomial $P(X)$ constructed from \Cref{lem:FrastfurtA3Euros} using the remaining set of equations.
 \label{lem:JoanEsGravissim}
\end{lem}

Note that generically, we will have $n$ complex solutions, up to $n!$ permutations. However, there is the possibility that some of the solutions lie at infinity in some pathological cases. Nevertheless, these cases form a zero-measure set which can in practice be avoided by adding an $\varepsilon$-perturbation to $d_k$, in the same spirit as in \Cref{eq:vanHaal}.

It is now clear that the bottleneck is solving \Cref{eq:Guardiolo} in \Cref{lem:FrastfurtA3Euros}. We propose two approaches in order to do so: In this section, a variant of the Faddeev-Leverrier algorithm to solve Newton's identities in order to find the roots power-sum symmetric polynomials. In \Cref{app:Amunike}, we propose a step-by-step computation of the solutions via Gr\"obner basis which could provide solutions for more general systems of equations.

We make two observations in order to solve \Cref{eq:Guardiolo}. The first one is that, if we consider a matrix $A$ with eigenvalues $\{x_1, \ldots, x_n\}$, then \Cref{eq:Guardiolo} can be thought of as
\begin{equation}
 \mathrm{Tr}[A^k] = z_k, \qquad 1\leq k \leq n.
 \label{eq:PitoVilanova}
\end{equation}
The second observation is that there exists a way to express the characteristic polynomial of a matrix $A$ in terms of $\mathrm{Tr}[A^k]$. Indeed, if $P(X) = \det{(X\mathbbm{1}-A)} =  (X-x_1) \cdots (X-x_n) = \sum_{k=0}^n c_k X^k$ is the characteristic polynomial of $A$, then $P(A) = 0$, and taking the trace in both sides yields such an equation. The Faddeev-Leverrier algorithm provides an easy to compute form for the coefficients $c_k$: They are given by
\begin{equation}
 c_{n-m} = \frac{(-1)^m}{m!}\left|
 \begin{array}{ccccc}
  \mathrm{Tr}[A] & m-1 & 0 & \cdots & 0\\
  \mathrm{Tr}[A^2] & \mathrm{Tr}[A] & m-2 &\cdots  & \\
  \vdots &  \vdots &  &  & \vdots\\
  \mathrm{Tr}[A^{m-1}] & \mathrm{Tr}[A^{m-2}] &\cdots & \mathrm{Tr[A]}&1\\
  \mathrm{Tr}[A^m] & \mathrm{Tr}[A^{m-1}] &\cdots &\cdots &\mathrm{Tr[A]}
 \end{array}
 \right| = 
  \frac{(-1)^m}{m!}\left|
 \begin{array}{ccccc}
  z_1 & m-1 & 0 & \cdots & 0\\
  z_2 & z_1 & m-2 &\cdots  & \\
  \vdots &  \vdots &  &  & \vdots\\
  z_{m-1} & z_{m-2} &\cdots & z_1&1\\
  z_{m} &z_{m-1} &\cdots &\cdots &z_1
 \end{array}
 \right|.
 \label{eq:Figo}
\end{equation}

The values of $x$ are found by finding all the roots of the polynomial $P(X) =  \sum_{k=0}^n c_k X^k$ with $c_n = 1$. In \Cref{app:Amunike} we give an alternative approach to solve \Cref{eq:Guardiolo} from a more algebraic point of view that gives more insight to the cominatorial structure underlying \Cref{eq:Guardiolo}.

Having an efficient way to represent a state of the form \Cref{eq:AnsuFati}, we can now use its MPS form to efficiently compute the fidelity of DMRG solutions, which are already given in the MPS formalism, for large $n$, thus being able to benchmark our method.

\subsection{Determining which symmetric states cannot be self-tested from their marginals}
\label{sec:SelfTest}

Self-testing is one of the most stringent protocols in the paradigm of device-independent quantum information processing. Self-testing consists in inferring, solely from the statistics of a Bell experiment, which quantum states and measurements are being used and performed, respectively \cite{MayersYaoST2004, Supic2019, YangPRAR2013}. Much of the existing work has been centered around the bipartite case \cite{BampsPRA2015, KaniewskiPRL2016}, partly motivated by its more accessible physical implmentations \cite{WangScience2018, ZhangNPJ2019}, but also motivated by its more accessible theoretical analysis, exploiting in most cases properties of the maximally entangled state of two qudits \cite{SATWAP, Kaniewski2018, SupicNJP2016}. In the multipartite case, the analysis becomes more complicated, although some ideas for the bipartite inspired some extensions \cite{SupicNJP2018}. In the multipartite case, symmetric states constitute a natural candidate to begin their study: for instance, the robust self-testing of the W state ($\ket{D_1}$) \cite{WuPRA2014} inspired schemes to self-test Dicke states of the form $\ket{D_k}$ \cite{Fadel2017, PhDWu, SupicNJP2018}. Nevertheless, these schemes use full-body correlators and require individual addressing, thus being less appealing from an experimental point of view. Therefore, some studies have been carried to find out whether self-testing is possible using only marginal information \cite{LiPRA2018} (see also \cite{BASTA, Augusiak2019}): In \cite{LiPRA2018}, some efforts showed that the three-qubit states maximally violating some of the translationally invariant, two-body Bell inequalities from \cite{TIPaper} could be self-tested using two-body correlators, thus giving a positive answer to this question.

Interestingly, the question of how much information from the statistics is needed (i.e., how many parties can one trace out) in order to self-test a quantum state is still open.
In this section, we aim at showing how our method can be used to guarantee a negative answer to the previous question 
and to give evidence towards a positive answer as well, depending on the uniqueness of the solution to \Cref{eq:AlbertJorquera}. More precisely, we show how our method can be used to determine which symmetric states could potentially be self-tested from marginals and which symmetric states definitely could not, because their marginals do not have a unique (modulo local unitaries) extension in the symmetric space.

Let us first consider an $n$-qubit density matrix being a projector onto a $n$-qubit Dicke state $\ket{D_k}$ as in \Cref{eq:Belletti}. We shall denote it $\rho_{n,k}$. Let $\sigma = \mathrm{Tr}_{1}(\rho)$ be the resulting density matrix from tracing out a single particle. In virtue of \Cref{eq:Deco}, we have
\begin{equation}
\sigma = {{n} \choose {k}}^{-1} \left( {{n-1} \choose {k-1}} \rho_{n-1,k-1}+ {{n-1} \choose {k}} \rho_{n-1,k} \right) \;.
\end{equation}
To gain some intuition, let us first study under which conditions is it possible to show that the purification of $\sigma$ is unique.
Following the spirit of Lemma 5.2 of \cite{ScaraniAPS2012}, we begin by considering a purification with an auxiliary system of the form
\begin{equation}
 \ket{\Phi} = \ket{D_{k-1}}\ket{P_1} + \ket{D_{k}}\ket{P_2},
\end{equation}
where the Dicke states act on $n-1$ qubits and
\begin{align}
\vert P_1\rangle &= \alpha_0 \vert 0\rangle \vert x_{10}\rangle + \alpha_1 \vert 1\rangle \vert x_{11}\rangle \\
\vert P_2\rangle &= \beta_0 \vert 0\rangle \vert x_{20}\rangle + \beta_1 \vert 1\rangle \vert x_{21}\rangle \;.
\end{align}
It follows from elementary algebra that the $(n-1)$-body RDM of $\vert\Phi\rangle\langle\Phi\vert$ is equal to $\sigma$ if, and only if,
\begin{equation}
\alpha_0 = 0 \quad,\quad \alpha_1 = \sqrt{\dfrac{k}{n}} \quad,\quad \beta_0 =  \sqrt{\dfrac{n-k}{n}}  \quad,\quad \beta_1 = 0 \;.
\end{equation}
Hence, in this case there exists a purification, it is unique, and it must be of the form
\begin{equation}
\vert \Phi\rangle = \vert D_{N}^{k}\rangle \vert x_{11} \rangle \;.
\end{equation}
\begin{cor}
 The $(n-1)-$partite reduced state of $\ket{D_k}$ uniquely determines $\ket{D_k}$ in the symmetric space.
 \label{cor:Fabregas}
\end{cor}

\begin{figure}[h!]
\centering
\begin{minipage}{.48\textwidth}
  \centering
  \includegraphics[width=0.8\linewidth]{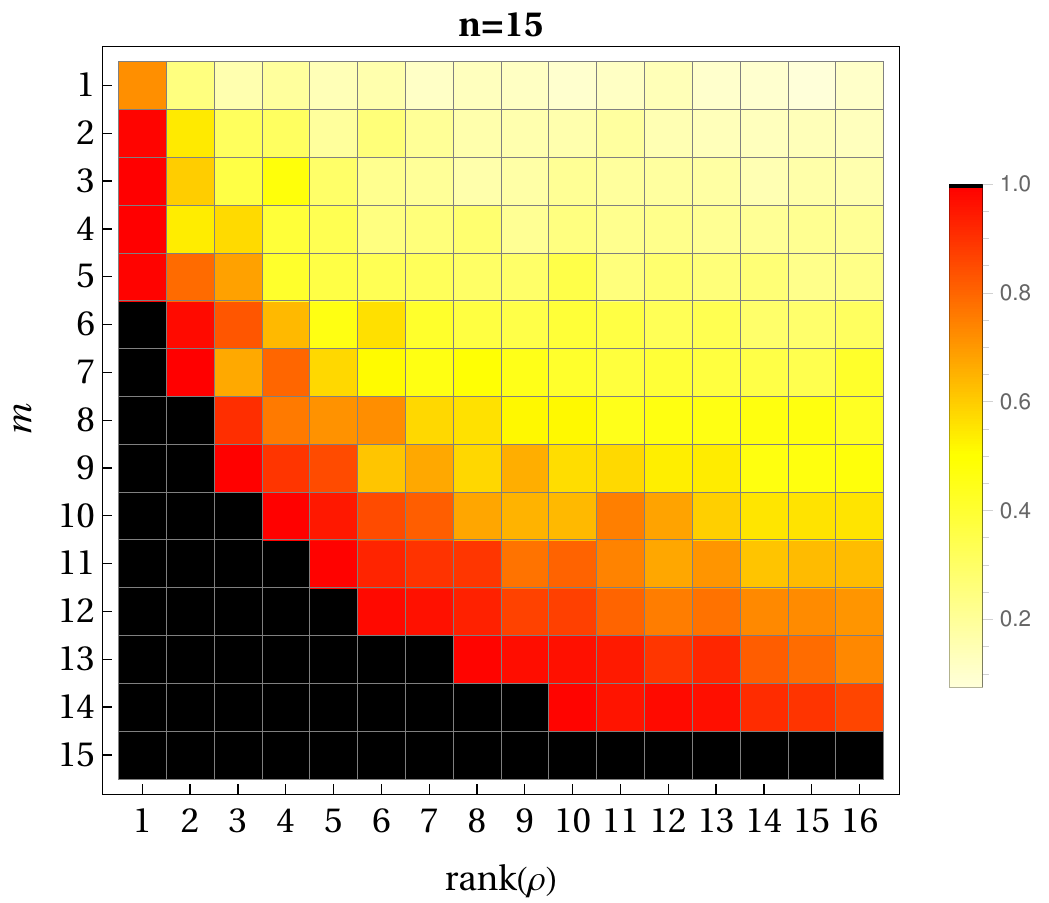}
\end{minipage}
\begin{minipage}{0.48\textwidth}
  \centering
  \includegraphics[width=0.8\linewidth]{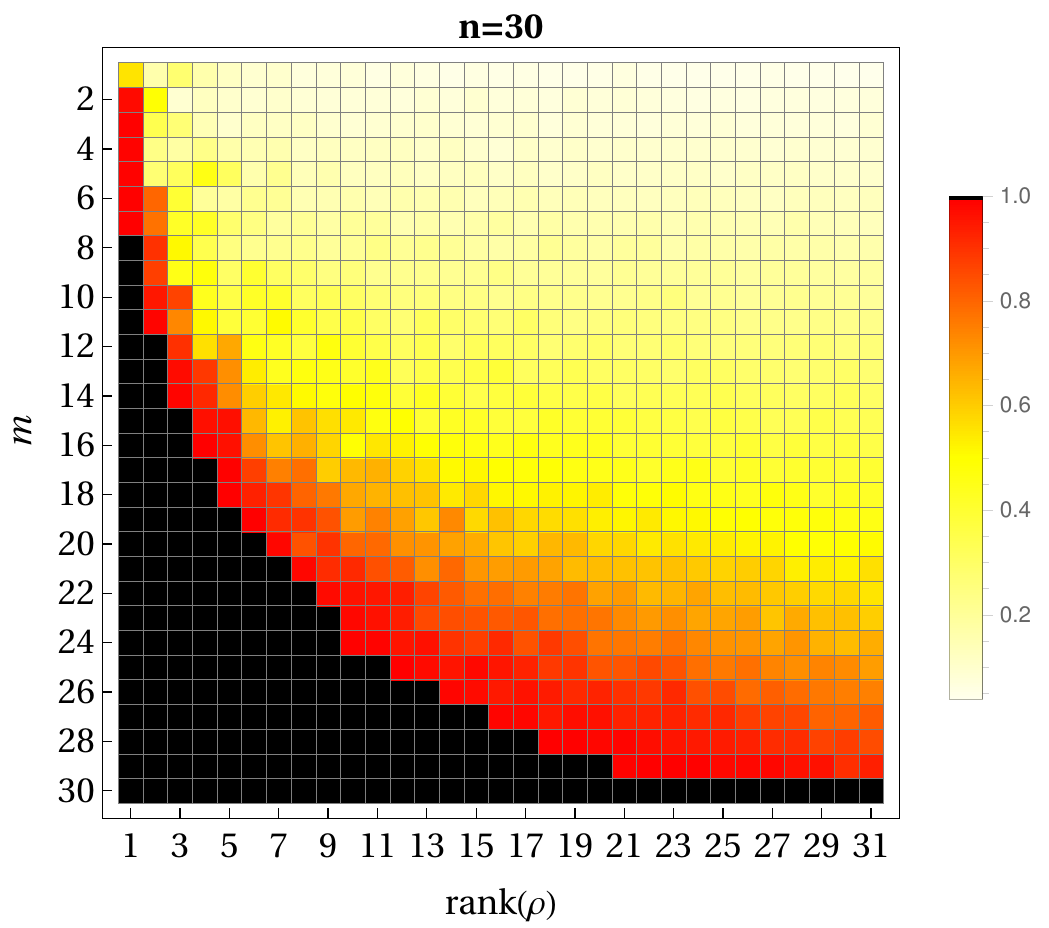}
\end{minipage}
\caption{Numerical results on the dependence between $n,m,\text{rank}(\rho)$ for an $m$-RDM to have a unique symmetric extension $\rho$ of $n=15$ and $n=30$ qubits respectively with a given $\text{rank}(\rho)$. For each case we have carried out 100 trials forcing the SDP to explore the feasible set in a random direction $A$ at each trial. The black squares correspond to the configurations for which the recovered global symmetric state has fidelity $>0.9999$ with the original global state for $100\%$ of the trials, thus providing evidence of having a unique symmetric extension. The numerical tolerance has been set to take into account the imprecision of the SDP solver. For the non-black squres, some of the trials have exhibited a fidelity $<0.9999$. For those cases, we show the minimal fidelity obtained out of all the trials as a way to illustrate the tolerance. One clearly observes a certain correlation between size of the RDM $m$ and $\text{rank}(\rho)$, showing more chances to have a unique extension for low $\text{rank}(\rho)$ by tracing out few particles.}
\label{fig:AFTconjecture}
\end{figure}

In \Cref{App:LP} we show how the above example can be generalized to tracing out any number of parties. The uniqueness of the extension is in one-to-one correspondence to the uniqueness of a linear program (see \Cref{eq:BeneficiEconomic}). We have numerically observed that such a solution is unique if we trace out up to $n-2$ parties for a basis Dicke state. However, it is not \textit{a priori} clear how generic the above property is. A more in-depth study suggests that generically, the uniqueness property depends on both the rank of the global density matrix and the number of parties traced out.
In \Cref{fig:AFTconjecture} we provide numerical evidence that generically the uniqueness of the symmetric extension depends on the number of particles $n$, the number of parties in the RDMs $m$ and the rank of the global density matrix $\rho$. We have followed the procedure below:
\begin{enumerate}
\item Generate a random symmetric state whose density matrix has a given rank
\item Use the compatibility conditions to obtain its RDM
\item Choose a random Hermitian matrix $A$ and find a new global state compatible with the RDM by making use of the SDP in \Cref{eq:AlbertJorquera}, but with objective function $-\langle A, \rho \rangle$ and check its fidelity with the original global state. The matrix $A$ forces the SDP to explore the feasible set in the direction given by $A$
\item We repeat step 3) a sufficient number of times (in our case, $100$ times).
If the fidelity remains always one up to numerical accuracy error, this is strong numerical evidence that the global state is unique. On the other hand, if the fidelity falls below one in some case, this indicates that too many parties have been traced out and the RDM would no longer be sufficient to self-test the original state, since it does not have a unique global extention of size $n$
\end{enumerate}

We note that, although mixed states in their generality cannot be self-tested, the fact that for some rank configurations and sizes of the RDM the extension to the symmetric state seems to be unique could open the door to a weaker form of self-testing, under the assumption that the global state is symmetric.
 
\section{Conclusions and outlook}
\label{sec:Concl}

In the present work we have presented a study of the QMP restricted to symmetric states. We have provided analytical compatibility conditions for an $m$-qudit RDM $\sigma$ to be compatible with an $n$-qudit global symmetric state $\rho$. We then use said compatibility conditions to answer the question of whether a given reduced density matrix $\sigma$ is compatible with a global symmetric state $\rho$ by turning it into a feasibility problem efficiently solvable via an SDP.
Our results have implications in different fields. We have explored some of them in several case-studies:
\begin{itemize}
\item We have developed a computationally efficient variational optimization method to upper bound the ground state energy of any local Hamiltonian. This method considers the resulting marginals to be compatible with a global symmetric state in order to carry out the optimization by means of SDP with the compatibility conditions as constraints. In order to benchmark the VM, we have considered several paradigmatic Hamiltonian spin models, that go from long-range to nearest-neighbor interactions. In general, we observe that the VM provides a good upper bound for ferromagnetic and long-range interactions, yielding exact results in the infinite-range limit; while it misses to capture antiferromagnetic short-range interactions, where the ground state has poor overlap with the symmetric space.
We have also used the compatibility conditions in order to obtain the half-system entanglement entropy in the symmetric space, which is a insightful quantity for many-body systems. Remarkably, we have observed that for some cases our variational method can also be used to approximately locate phase transitions. The reason for that lies that the change of properties of the ground state in a phase transition also maifest, to some extent, in the symmetric space projection and are therefore captured by our method. Another observed feature is that for some specific models the VM has recovered the first excited state, instead of the ground state, in some regions of the phase diagram. Finally, we have observed a significant speed advantage of the VM compared to a typical DMRG algorithm. The advantage of the VM lies on the low memory storage required and high speed, making the VM a suitable candidate for a first order exploration of large sets of parameters characterizing the phase diagram of spin Hamiltonians. While we have only considered qubits, qutrits and chain configurations, our VM is straightforwardly applicable to any qudit and lattices of arbitrary geometry and dimension. We leave open to implement and explore the VM in corresponding cases of interest.

\item We have considered the VM in the context of Bell non-local correlations. In particular, we have explored its synergy with the so-called 2-body permutationally invariant Bell inequalities. The results in this context are two-fold: First, we have shown how the variational method comes as a natural tool to optimize a multipartite 2-body PIBI in order to find whether the inequality detects non-local correlations; Second, we have used the low computational cost of the variational method to look for non-local correlations in a spin-$1/2$ XXZ chain under a transverse field, narrowing the parameters to be considered and eventually leading to the detection of non-local correlations the ground state with $n=128$ parties. We have also considered another spin-$1/2$ XXZ chain this time with periodic boundary conditions with longitudinal magnetic field, detecting non-local correlations on its ground state and first excited state of a specific phase. The tool we have here presented can be readily used in the context of Bell correlation depth \cite{BaccariPRA2019} or DI entanglement depth certification \cite{AloyPRL2019, TuraPRA2019} in the context of 2-body PIBIs.

\item We have developed an analytical methodology to derive a translationally invariant diagonal matrix-product state representation of bond dimension $n$ for pure symmetric states. This result is generic, and could be of independent interest. For our purposes, we have used it to transform the symmetric state solution obtained with the VM into a translationally invariant, diagonal, matrix product state, allowing us to check its fidelity with the DMRG solution.

\item Finally, we have shown how the compatibility conditions can be used to determine which symmetric states $\rho$ cannot be self-tested solely from their marginals. Remarkably, we present numerical evidence suggesting a correlation between the size of the global state $n$, its rank $\text{rank}(\rho)$, how many particles $m$ remain in the observed RDM and the uniqueness of a symmetric global state. This uniqueness property could open the way to a weaker form of self-testing, that usese the assumption that the global state is symmetric.
\end{itemize}

One may wish to impose the condition of the global state being pure; i.e., $\rho$ being a rank-$1$ projector. Although this condition clearly breaks the convexity of the SDP program, one can combine the VM with an iterative projection onto the principal component of $\rho$ \cite{TIPaper, HuberJPA2018} in order to converge to a rank-$1$ solution. Furthermore, one may also wish to explore the role of different symmetries in the SDP. Whether there exists a SDP invariant formulation of our problem \cite{BachocHSCPO2011, TavakoliPRL2019} that could allow it to be formulated for other symmetry groups is unclear and we leave it for future research.
In a following work, we shall investigate a variations of the VM in order to perform tomography/fidelity estimates with respect to a target symmetric state. This is of wide experimental relevance, as in the case of Bose-Einsten condensate where only partial information (e.g. not an informationally complete set of measurements) is available \cite{SchmiedScience2016, FadelScience2018}.

\section*{Acknowledgments}
We thank A. Ac{\'i}n, F. Huber and M. Sanz for inspiring discussions. M.F. acknowledges support from the Swiss National Science Foundation. J.T. thanks the Alexander von Humboldt foundation for support. We acknowledge the Spanish Ministry MINECO (National Plan 15 Grants No. FISICATEAMO, No. FIS2016-79508-P, No. SEVERO OCHOA No. SEV-2015-0522, FPI), European Social Fund, Fundació Cellex, Generalitat de Catalunya (AGAUR Grant No. 2017 SGR1341 and CERCA Programme), ERC AdG OSYRIS and NOQIA, and the National Science Centre, Poland-Symfonia Grant No. 2016/20/W/ST4/00314.

\appendix

\section{Alternative solution to the system of equations in \Cref{lem:FrastfurtA3Euros}}
\label{app:Amunike}

Before introducing the form of $P(X)$ let us motivate its definition by illustrating the idea with a sequence of examples. In these examples, we turn \Cref{eq:Guardiolo} into an equivalent system that is much easier to solve. In algebraic geometry terms, the second system forms a reduced Groebner basis, meaning that its first equation is a polynomial in a single variable, the second is a polynomial in the previous variable and a new one, etc. This allows one to find all the solutions by solving only univariate polynomials and plugging the found roots into the next equations by substitution.
\begin{itemize}
 \item $N=2$. Solving the system of equations
 \begin{equation}
 \left\{
  \begin{array}{ccc}
   x_1 + x_2 -z_1&=& 0\\
   x_1^2 + x_2^2 -z_2&=& 0
  \end{array}
   \right.
  \label{eq:343}
 \end{equation}
is equivalent to solving
\begin{equation}
 \left\{
  \begin{array}{ccc}
   2x_2^2 -2z_1 x_2 + (z_1^2-z_2) &=& 0\\
   x_1 + x_2 -z_1&=& 0
  \end{array}
   \right.
  \label{eq:343b}
\end{equation}
 \item $N=3$. Solving the system of equations
 \begin{equation}
 \left\{
  \begin{array}{ccc}
   x_1 + x_2 +x_3-z_1&=& 0\\
   x_1^2 + x_2^2 +x_3^2-z_2&=& 0\\
   x_1^3 + x_2^3 +x_3^3-z_3&=& 0
  \end{array}
   \right.
  \label{eq:4231}
 \end{equation}
is equivalent to solving
\begin{equation}
 \left\{
  \begin{array}{ccc}
   6x_3^3 - 6z_1x_3^2+3(z_1^2-z_2)x_3 + (-z_1^3+3z_1z_2 -2z_3) &=& 0\\
   2x_3^2-2(z_1-x_2)x_3 + [(z_1-x_2)^2-(z_2-x_2^2)]&=&0\\
   x_2 + x_3 - (z_1-x_1)&=& 0
  \end{array}
   \right.
  \label{eq:4231b}
\end{equation}
 \item $N=4$. Solving the system of equations
 \begin{equation}
 \left\{
  \begin{array}{ccc}
   x_1 + x_2 +x_3+x_4-z_1&=& 0\\
   x_1^2 + x_2^2 +x_3^2+x_4^2-z_2&=& 0\\
   x_1^3 + x_2^3 +x_3^3+x_4^3-z_3&=& 0\\
   x_1^4 + x_2^4 +x_3^4+x_4^4-z_4&=& 0
  \end{array}
   \right.
  \label{eq:541}
 \end{equation}
is equivalent to solving
\begin{equation}
 \left\{
  \begin{array}{ccc}
  24x_4^4 - 24 z_1 x_4^3 + 6(z_1^2-z_2)x_4^2-2(z_1^3-3z_1z_2+2z_3)x_4 + (z_1^4 -6z_1^2z_2 +3z_2^2+8z_1z_3-6z_4)&=&0\\  
   6x_4^3 -6(z_1-x_3)x_4^2 +3([z_1-x_3]^2-[z_2-x_3^2])x_4 - ([z_1-x_3]^3 - 3[z_1-x_3][z_2-x_3^2] +2[z_3-x_3^3]) &=& 0\\
   2x_4^2-2(z_1-x_2-x_3)x_4 + [(z_1-x_2-x_3)^2-(z_2-x_2^2-x_3^2)]&=&0\\
   x_2+x_3 + x_4 - (z_1-x_1)&=& 0
  \end{array}
   \right.
  \label{eq:541b}
\end{equation}
\end{itemize}

From the above examples the recursion is clear. In the \emph{easy} systems \Cref{eq:343b}, \Cref{eq:4231b}, \Cref{eq:541b} the first equation is a polynomial in a single variable $x_n$. The rest of the equations correspond to the system of equations for $n-1$ with a slight transformation, where we have decreased by $1$ the index of $x_i$; i.e., $x_i \mapsto x_{i-1}$ and we have made the substitution $z_i \mapsto z_i-x_{n-1}^i$ in the first equation, $z_i \mapsto z_i-x_{n-2}^i$ in the second equation and so on until we substitute $z_1 \mapsto z_1 - x_1$ in the last one. Note that since the first equation is a polynomial in $x_n$, the second equation is a polynomial in $x_n, x_{n-1}$, the third equation a polynomial in $x_n, x_{n-1}, x_{n-2}$ and so on the transformed systems form a reduced Groebner basis and are therefore easy to solve.

Before thinking of writing the Groebner basis in its full generality, let us observe the following:
\begin{cor}
 Let $P(X)$ be the first element of the Groebner basis for \Cref{eq:Guardiolo} (i.e., the left hand side of the first equation in the systems \Cref{eq:343b}, \Cref{eq:4231b}, \Cref{eq:541b}, etc.). Since the system of equations \Cref{eq:Guardiolo} is permutationally invariant, we must have
 \begin{equation}
  P(X) = (X-x_1)\cdots (X-x_n),
 \end{equation}
 i.e., the roots of $P$ correspond to the values of $x_i$, up to a permutation.
\end{cor}

Therefore, we only need to find the general form of $P(X)$. The coefficients of $P(X)$ are closely related to the partitions of $n$. Let us define the following:
\begin{defn}
 Let $\boldsymbol{\lambda}\vdash m$ denote a partition of $m$; i.e., $\boldsymbol{\lambda} = (\lambda_1^{\mu_1}, \ldots \lambda_k^{\mu_k})$ where $\sum_{i=1}^k \mu_i\lambda_i = m$ and $\lambda_i > \lambda_{i+1}$ with $\lambda_i, \mu_i \in \mathbbm{N}$.
 We define the polynomial 
 \begin{equation}
  Q_{m}(\boldsymbol{z}) := \sum_{\boldsymbol{\lambda}\vdash m}\xi_{\boldsymbol \lambda}\prod_{i=1}^kz_{\lambda_i}^{\mu_i},
  \label{eq:CoCu}
 \end{equation}
 where
 \begin{equation}
  \xi_{\boldsymbol{\lambda}} = m! \prod_{i=1}^k \frac{(-1)^{\mu_i}}{\mu_i! \lambda_i^{\mu_k}}.
 \end{equation}
 We define by convention $Q_0 := 1$.
\end{defn}

Note that $\sum_{\boldsymbol{\lambda}\vdash m} |\xi_{\boldsymbol{\lambda}}| = m!$ since $\xi_{\boldsymbol{\lambda}}$ counts (with sign) the number of permutations of $m$ elements of cycle type $\boldsymbol{\lambda}$.
In addition, we remark that the number of partitions $p(m)$ of a given integer $m$ scales as $\log p(m) \sim C \sqrt{m}$, where $C$ is a universal constant. This makes the sum \Cref{eq:Cocu} prohibitive to evaluate already for modestly large values of $m$. However, as shown in \Cref{sec:MPS}, it is possible to efficiently compute $Q_{m}(\boldsymbol{z})$ without splitting it into its different summands.

\begin{defn}
 We define $P(X)$ to be
 \begin{equation}
  P(X) := \sum_{m=0}^n \frac{n!}{m!}Q_m(\boldsymbol{z})X^{n-m}.
  \label{eq:Pedrito}
 \end{equation}
\end{defn}

Now that we know how to obtain the $\boldsymbol{x}$ that satisfy $\boldsymbol{z}$ in \Cref{eq:Guardiolo}, let us turn to the system of equations that actually arises from \Cref{eq:umtiti}. Note that \Cref{eq:Guardiolo} does not take into consideration the $z_0$ term, but we incorporating the condition that $A_0 \propto \mathbbm{1}$. 
Then the system of equations of interest becomes \Cref{eq:LaportaPresidentCatalunyaIndependent}.

The system of equations \Cref{eq:Guardiolo} is also known the the power sum ideal. Its reduced Groebner Basis is found as the elimination ideal of the power sums.
\begin{cor}
 The elimination ideal of the power sums gives the compatibility conditions on the weights $d_k$ of \Cref{eq:AnsuFati} to be representable with a diagonal TI MPS of bond dimension $D < n$.
\end{cor}
Indeed, let us consider $n=4$ and $D=3$. The elimination ideal of the power sums of three variables and degree four is
\begin{equation}
 \langle x_1 + x_2 + x_3 - z_1, x_1^2 + x_2^2 + x_3^2 - z_2, x_1^3 + x_2^3 + x_3^3 - z_3, x_1^4 + x_2^4 + x_3^4 - z_4\rangle \cap \mathbbm{K}[z_1,z_2,z_3,z_4] = \langle q(z_1,z_2,z_3,z_4)\rangle,
 \label{eq:MagicAndreu}
\end{equation}
where
\begin{equation}
 q(z_1,z_2,z_3,z_4) = z_1^4 - 6z_1^2z_2 + 3 z_2^2 + 8 z_1z_3 - 6z_4
 \label{eq:AndreuMagic}
\end{equation}
Note that the compatibility polynomial in \Cref{eq:AndreuMagic} is precisely the same polynomial $Q_4(\boldsymbol{z})$ in the constant term of the univariate polynomial in \Cref{eq:541b}.
Hence, all the symmetric Dicke states for which the $\boldsymbol{z}$ obtained from their $d_k$ belongs to the elimination ideal of the power sums of $D$ variables with degree $n$ are representable as a diagonal TI MPS of the form $A_0 \propto \mathbbm{1}$ and $A_1 = \mathrm{diag}(\boldsymbol{x})$.

\section{Linear programming approach for Dicke-diagonal states}
\label{App:LP}

Let us see how we can now apply \Cref{eq:VictorValdes} in a more systematic way to determine that the states of the Dicke basis are the only ones in which \Cref{cor:Fabregas} applies.

Let us consider $\rho$ as a rank-$1$ projector onto a quantum state of the form \Cref{eq:AnsuFati}. In virtue of \Cref{eq:Cocu} we have (note that for qubits the partition of $n$ is identified by a single number, therefore we write $\alpha$ instead of $\boldsymbol{\alpha}$)
\begin{equation}
 \sigma^{\alpha}_\beta = \sum_{p=0}^{n-m}\sqrt{\frac{{m \choose \alpha}{m \choose \beta}}{{n \choose \alpha + p}{n \choose \beta + p}}}{n - m \choose p} d_{\alpha + p}^* d_{\beta + p}.
 \label{eq:QQ}
\end{equation}
\begin{itemize}
 \item If we set $d_{\alpha} = \delta(\alpha - k)$ then we have
 \begin{equation}
  \sigma^{\alpha}_\beta = \delta(\alpha - \beta){n \choose k}^{-1}{m \choose \alpha} {n-m \choose k-\alpha} I_{[0,n-m]}(k-\alpha),
 \end{equation}
where $I_{S}(x)$ is the indicator function, which evaluates to $1$ if $x \in S$ and $0$ otherwise. This allows us to write the set of equations for any basis Dicke state:
\begin{equation}
 \sum_{p=0}^{n-m}\rho^{\alpha + p}_{\beta + p} \sqrt{\frac{{m \choose \alpha}{m \choose \beta}}{{n \choose \alpha + p}{n \choose \beta + p}}}{n-m \choose p} =  \delta(\alpha - \beta){n \choose k}^{-1}{m \choose \alpha} {n-m \choose k-\alpha} I_{[0,n-m]}(k-\alpha).
 \label{eq:FrancescSatorras}
\end{equation}
 \begin{itemize}
  \item If we take $m=n-1$, we recover the result of \Cref{cor:Fabregas} in the following way: Since
\begin{equation}
 \sigma = {n \choose k}^{-1}\sum_{\alpha = k-1}^k {n-1 \choose \alpha} \ket{\alpha}\bra{\alpha},
\end{equation}
the conditions of the SdP \Cref{eq:VictorValdes} can be now rewritten as
\begin{equation}
 \sum_{p=0}^1 \rho^{\alpha + p}_{\beta + p} \sqrt{\frac{{n-1\choose \alpha}{n-1 \choose \beta}}{{n \choose \alpha + p}{n \choose \beta + p}}}{1 \choose p}= \delta(\alpha - \beta){n-1 \choose \alpha}{n \choose k}^{-1} I_{[0,1]}(k-\alpha).
 \label{eq:FrankDeBoer}
\end{equation}
We note that the right hand side of \Cref{eq:FrankDeBoer} is zero if $\alpha > k$ or $\alpha < k-1$. In these cases, in the diagonal ($\alpha = \beta$) we have a condition of the form
\begin{equation}
 \rho^{\alpha}_{\alpha} \xi_{\alpha} + \rho^{\alpha+1}_{ \alpha + 1} \xi_{\alpha+1} = 0,
 \label{eq:DeJong}
\end{equation}
for some $\xi_{\alpha} > 0$ that we do not need to write here explicitly. Now, the semidefinite positivity condition on $\rho$ from the SdP \Cref{eq:VictorValdes} implies that the diagonal elements must be non-negative: $\rho^{\alpha}_{\alpha}\geq 0$. Hence, for all $\alpha \geq k+1$ and $\alpha \leq k-2$ we must have $\rho^{\alpha}_{\alpha} = \rho^{\alpha+1}_{\alpha+1} = 0$. The condition $\rho \succeq 0$ further implies that all the elements in the respective rows and columns must be zero. Therefore, the only non-zero element $\rho^{\alpha}_{\alpha}$ left is $\rho^{k}_{k}$, which must be $1$ in virtue of \Cref{eq:FrankDeBoer}.
\item If we trace out two parties, i.e., we take $m=n-2$, then a similar argument follows: We see that for $\alpha > k$ or $\alpha < k-2$ we have a condition in the diagonal similar to the form of \Cref{eq:DeJong}
\begin{equation}
 \rho^{\alpha}_{\alpha} \xi_{\alpha} + \rho^{\alpha+1}_{ \alpha + 1} \xi_{\alpha+1} + \rho^{\alpha+2}_{ \alpha + 2} \xi_{\alpha+2}= 0.
 \label{eq:DeLiejs}
\end{equation}
Again, in \Cref{eq:DeLiejs} we have a linear combination of $\rho^{\alpha + p}_{\alpha + p} \geq 0$ (because $\rho \succeq 0$) with strictly positive weights $\xi_{\alpha + p}^{\alpha+ p} > 0$. This implies that $\rho^{\alpha}_{\alpha} = 0$ for $\alpha \neq k$, and a similar argument follows. However, one needs to be careful in counting the number of zero and non-zero equations: we need $n>4$ for \Cref{eq:DeLiejs} to exist. To this end, let us see the general case:
\item If we trace out $n-m$ parties, then we generalize the last two points: For $\alpha > k$ or $\alpha < k-(n-m)$ the condition on the diagonal is
\begin{equation}
 \sum_{p=0}^{n-m} \rho^{\alpha + p}_{\alpha + p} \xi_{\alpha + p} = 0,
 \label{eq:VanGal}
\end{equation}
which implies $\rho^{k+1+p}_{k+1+p} = \rho_{k-1-p}^{k-1-p} = 0$ for $p \geq 0$. This condition is nontrivial as long as the number of equations ($m-1$) is greater than the number of nonzero left hand sides ($n-m+1$), i.e., whenever $m>n/2$.
Therefore, the condition $\rho \succeq 0$ implies that all the off-diagonal elements must be zero and therefore $\rho^k_k = 1$.
 \item Suppose we trace out $n-m$ parties. Then we have the following system of equations:
 \begin{align}
  \left(
  \begin{array}{cccccc}
   {n-m\choose 0}&{n-m \choose 1}& \cdots & {n-m \choose n-m}&\cdots & 0\\
   0&{n-m\choose 0}&\cdots&{n-m \choose n-m-1}&\cdots & 0\\
   \vdots&\vdots&\ddots&\vdots&\ddots&\vdots\\
   0&0&\cdots&{n-m \choose n -2m}&\cdots & {n-m \choose n-m}
  \end{array}
  \right)
  \left(
  \begin{array}{c}
   x_0\\x_1\\\vdots\\x_{n-m}\\\vdots \\x_n
  \end{array}
  \right)
  =
  \left(
  \begin{array}{c}
   {n-m\choose k} I_{[0,n-m]}(k)\\
   {n-m\choose k-1} I_{[0,n-m]}(k-1)\\
   \vdots\\
   {n-m\choose k-m} I_{[0,n-m]}(k-m)
  \end{array}
  \right),
  \label{eq:BeneficiEconomic}
 \end{align}
where we have defined for simplicity $x_p := \rho^{p}_p{n\choose k}/{n \choose p}$. If $m > n/2$, there must necessarily be zeroes in the right hand side of \Cref{eq:BeneficiEconomic}.
 \end{itemize}
\end{itemize}

The question about uniqueness of solutions of linear programs \cite{MangasarianBook1984} and semidefinite programs \cite{ZhuIEEE2010, AlfakihDAM2007} is an intensive field of research, due to its connection to rigidity theory. For instance, the general solution to the uniqueness of \Cref{eq:AlbertJorquera} can be expressed via
\begin{thm} \cite{ZhuIEEE2010}
If $\rho$ is a max-rank solution of \Cref{eq:AlbertJorquera}, and we write $\rho = L^\dagger L$, where $L \in \mathbbm{C}^{r\times n}$, then $\rho$ is the unique solution of \Cref{eq:AlbertJorquera} if, and only if, the kernel of the linear space spanned by $L^\dagger A^{\boldsymbol{\alpha}}_{\boldsymbol \beta} L$ is trivial.
\end{thm}
\begin{cor} \cite{ZhuIEEE2010}
 If all the solutions to \Cref{eq:AlbertJorquera} share the same rank, then the solution must be unique.
\end{cor}

\bibliography{mylib}

\end{document}